\pdfoutput=1
\documentclass[12pt]{article}

\usepackage{amsmath,amssymb,amsbsy,amsfonts,latexsym,graphicx}
\usepackage{hyperref}
\usepackage{color,array,subfigure}
\usepackage{cite}

     \renewcommand{\L}{\Lambda}

\allowdisplaybreaks

\evensidemargin \oddsidemargin

 \def\L{{\Lambda}}

\begin{document}


\begin{titlepage}

\renewcommand{\thefootnote}{\fnsymbol{footnote}}



\vspace{15mm}
\baselineskip 9mm
\begin{center}
  {\Large \bf 
{  
Impurity Effect on Hysteric Magnetoconductance: Holographic Approach} 
 }
\end{center}

\baselineskip 6mm
\vspace{10mm}
\begin{center}
Kyung Kiu Kim$^{1,a}$, Keun-Young Kim$^{2,b}$, Sang-Jin Sin$^{3,c}$ and Yunseok Seo$^{4,d}$
 \\[10mm] 
  $^1${\sl Department of Physics and Astronomy, Sejong University, Seoul 05006, Korea}
   \\[3mm]
    $^2${\sl School of Physics and Chemistry, Gwangju Institute of Science and Technology,  \\Gwangju 61005, Korea}
    \\[3mm] 
$^3${\sl Department of Physics, Hanyang University, Seoul 04764, Korea}
    \\[3mm]      
    $^4${\sl College of General Education, Kookmin University, Seoul 02707, Korea}
    \\[3mm]
  {\tt  ${}^a$kimkyungkiu@sejong.ac.kr, ${}^b$fortoe@gist.ac.kr, ${}^c$sjsin@hanyang.ac.kr, ${}^d$yseo@kookmin.ac.kr}
\end{center}

\thispagestyle{empty}

\vspace{1cm}
\begin{center}
{\bf Abstract}
\end{center}
\noindent
In this paper we study a hysteric phase transition from weak localization  phase to hysteric magnetoconductance phase using gauge/gravity duality. This  hysteric phase is triggered by a spontaneous magnetization related to $\mathbb{Z}_2$ symmetry and time reversal symmetry in a 2+1 dimensional system with momentum relaxation. We derive thermoelectric conductivity formulas describing non-hysteric and hysteric phases. At low temperatures, this magnetoconductance shows similar phase transitions of topological insulator  surface states. We also obtain hysteresis curves of Seebeck coefficient and Nernst signal. It turns out that our impurity parameter changes magnetic properties of the dual system.  This is justified by showing increasing susceptibility and the spontaneous magnetization with increasing impurity parameter. 
\\ [15mm]

\vspace{5mm}
\end{titlepage}

\baselineskip 6.6mm
\renewcommand{\thefootnote}{\arabic{footnote}}
\setcounter{footnote}{0}

\section{Introduction and Summary}

Recently, topological insulators (TIs) have attracted considerable interest in theoretical research and potential applications. One of the most decisive factors that characterize TIs is the existence of gapless surface-state protected by time reversal symmetry(TRS). See \cite{HasanKane,Ando:2013bqa} for reviews. Since the underlying physics is sensitive to TRS and the band structure, it is important to adjust the band gap for application of TIs. So one may try to break the TRS to control TIs naturally. One way to achieve TRS breaking is doping TIs with magnetic impurities. Such a method has already been realized in numerous experiments, which provided reliable data for various materials \cite{TIexps-1,TIexps-2,Cr-dopedTI-1,Cr-dopedTI-2,Cr-dopedTI-3}.

Even though the bulk of a TI is an insulator, currents still flow on the surface of TIs. Therefore, the referred experimental data were obtained by measuring magnetoconductance under magnetic doping. These experiments show a crossover from weak anti-localization (WAL) to weak localization (WL) as the magnetic doping concentration increases and also there is a novel phase transition distinguished by hysteresis behavior of magnetoconductance. From now on, we will refer to this hysteric phase as HMC phase.

One of the interesting phenomena is that the hysteric behavior occurs at a certain magnetic impurity density. For example, Cr doped BiTe topological insulator shows this phenomenon \cite{TIexps-1,TIexps-2}. There is no hysteric behavior in the magnetoconductance before the doping parameter $x$ of Cr is greater than $x=0.14$.  However, when the doping parameter is greater than $x=0.14$, the spontaneous magnetization appears and hence the hysteric behavior of magnetoconductance begins to appear.  The details of this mechanism are still mysterious but several researchers proposed that this phenomenon comes from the strong correlation between the lattice and the magnetic order of the impurities.  When the interaction becomes strong, we cannot analyze the system in perturbative calculation, we need new method for it. While it is difficult to understand the hysteric behavior, there has been significant progress in the non-hysteric WAL and WL phases including the crossover, $e.g.$, \cite{WAL-WL} based on Hikami-Larkin-Nagaoka model \cite{Hikami_model}.  There are studies on magnetic properties of topological insulator using field theory methods \cite{YYYang,Zarezad}.

There is a holographic study about the crossover between WAL and WL phases\cite{Seo:2017oyh,Seo:2017yux} by adding Chern-Simons like coupling with impurity and gauge field. In theses works, we introduce two types of linear-axion fields, whose nomenclature will be explained later. One type of axion field interacts with gauge field via Chern-Simons term and the other one does not. Two types of axion fields can be interpreted as a magnetic and non-magnetic impurity. Following standard gauge/gravity technique, we calculate magnetoconductance and we find that the results well agree with the experimental data of the surface state of the TIs with magnetic impurity. However, the model does not show any hysteric behavior even for the large impurity density limit.

Of course it would be fantastic to find a model covering the crossover and the phase transition. However even the phase transition from WL phase to HMC phase has been less studied.
In this note we are devoted to constructing a simplest holographic model showing this phase transition. Holographic method is based on AdS/CFT correspondence \cite{Maldacena:1997re,Aharony:1999ti} which provides a powerful tool to study various strongly coupled field theories. Later, this method has been widely applied to various research areas. In particular, topics of condensed matter physics are considered as major applications of the correspondence, {\it e.g.} \cite{Hartnoll:2008vx,Hartnoll:2007ih,Hartnoll:2009sz,Hartnoll:2009ns}. Also, the magnetic property of holographic matters has been studied using various gravity models, {\it e.g.} \cite{Hartnoll:2007ip,Hartnoll:2007ih,Lindgren:2015lia,Lindgren:2015lia2}.

Recently, we constructed a four-dimensional gravity model describing spontaneous magnetization and we obtained hysteric magnetization curves of the dual 2+1 dimensional system \cite{Kim:2019lxb}. This model is suitable for describing magnetic property of two-dimensional surface material. In this paper we study an extended model by adding an  matter field $\psi^{\mathcal{I}}$ to the previous one. This field is the previously mentioned linear-axion field since it is linear in spatial boundary coordinate $x^i$ and breaks a shift symmetry. This linear-axion field is known to explain the effect of momentum relaxation from impurity or lattice, $e.g.$, \cite{Andrade:2013gsa,Gouteraux:2014hca,Donos:2014cya,Donos:2014uba,Kim:2014bza,Blake:2015ina,Kim:2015wba,Kim:2015dna,Kim:2016hzi}. 
Thus main goal of the present work is to show that the system undergoes the phase transition from WL phase to HMC phase by increasing this impurity. This phase transition is originated from $\mathbb{Z}_2$ symmetry breaking associated with a scalar operator condensation at low temperature. The detailed discussion on this $\mathbb{Z}_2$ symmetry breaking was provided in \cite{Kim:2019lxb}. Since the only difference of the present model from the previous study is consideration of the linear-axion field which has nothing to do with the $\mathbb{Z}_2$ symmetry, the discussion of \cite{Kim:2019lxb} can be applied to this extended case\footnote{In the presence of external magnetic field, it is important to take into account time reversal transformation for each field. We will give simple comment and future direction for this issue in conclusion.}.

\begin{figure}[] 
\begin{centering}
    {\includegraphics[width=7.2cm]{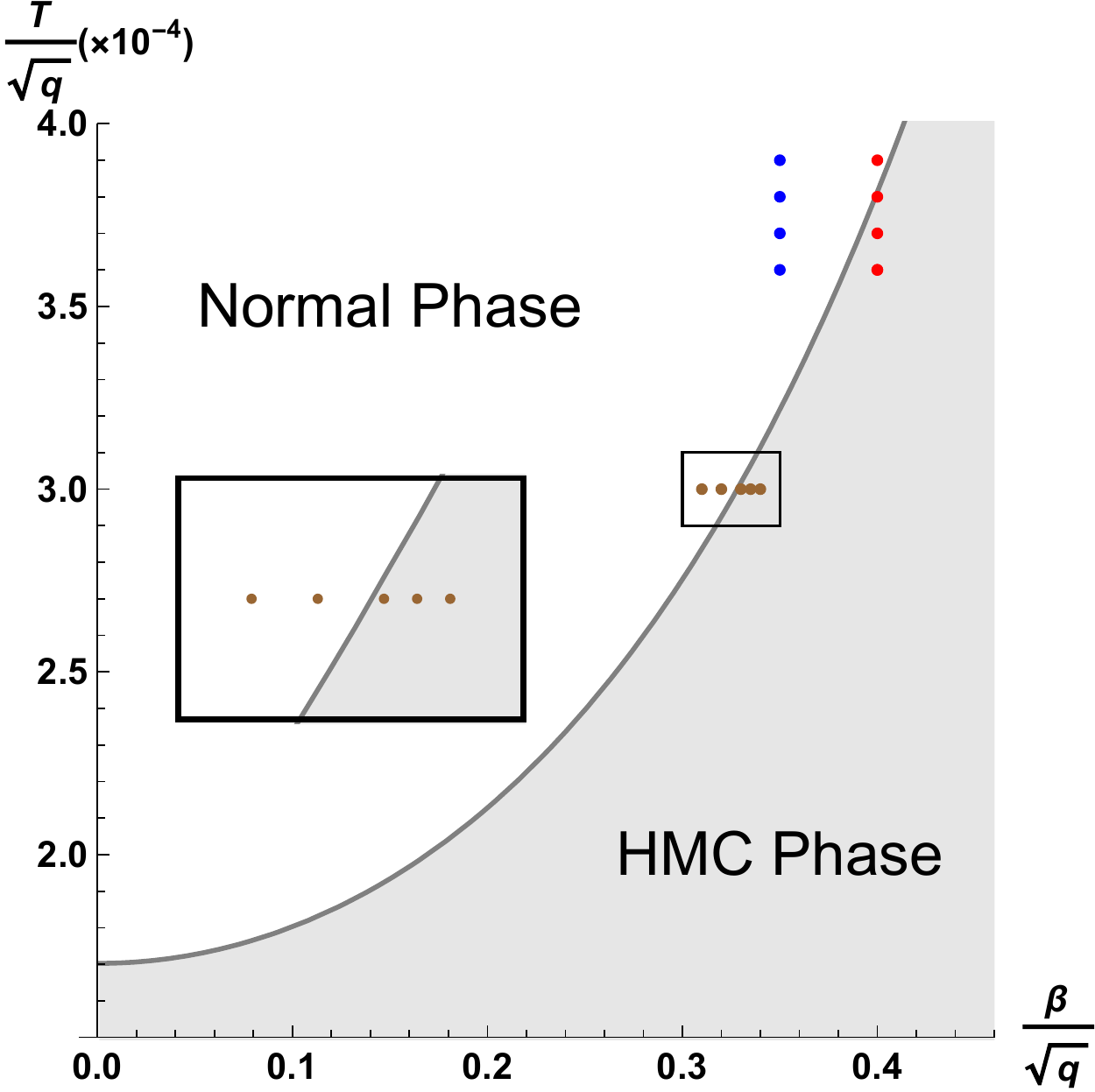}  }
       
    \caption{This shows the phase diagram near the zero temperature. The brown dots denote $T/\sqrt{q}=3\times 10^{-4}$. The corresponding transport coefficients are shown in Figure \ref{fig:beta_Hysteresis} and \ref{fig:beta_SN}. The blue and red dots represent $\beta/\sqrt{q}=0.35$ and $0.4$, respectively. 
Their transport coefficients are shown in Figure \ref{fig:T:conductivity} and \ref{fig:T:SN}. 
} \label{fig:Pdiagram}
\end{centering}
\end{figure}

Before discussing the phase transition, it is very important to clarify that this impurity dual to the linear-axion field can really be identified with the magnetic impurity. As one can see in (\ref{S_B}), the linear-axion field is not coupled to the magnetic field directly so it does not seem clear that this axion field carries magnetic moments in the dual system. However, the axion field can interact indirectly with the magnetic field through bulk back-reaction. In order to confirm this, we compute the spontaneous magnetization and susceptibility in the HMC and WL phases, respectively. The Figure \ref{fig:MvsBeta} shows that the susceptibility and the magnetization increase as the impurity density $\beta$ increases. In this figure we control only $\beta$ with fixed charge density without other sources, so it is justified that the impurity strength $\beta$ changes magnetic properties through nontrivial bulk back-reaction. Thus our model shed a light on holographic construction of magnetic impurity.

As we introduced earlier, the magnetic conductance is the observable which distinguishes the HMC phase from the non-hysteric WL phase in experiments, {\it e.g.} \cite{TIexps-1,TIexps-2}. Thus we derive the conductivity formulas, (\ref{conductivity}) and (\ref{alpha}), by using a holographic method. This holographic conductivity is usually given by horizon data of black branes. See \cite{Donos:2014cya,Blake:2015ina,Donos:2015bxe,Kim:2015wba} for examples of derivation for different models. Since the phase transition comes from the spontaneous magnetization, it is possible to construct the phase diagram in terms of the impurity density and temperature for a given charge density $q$ using a numerical method. The phase diagram is provided in Figure \ref{fig:Pdiagram}, where it is shown that larger $\beta$, describing more impurities, or lower temperature triggers the phase transition. This implies that the axion field which is non-magnetic itself enhances spontaneous condensation of scalar field which is magnetic\footnote{We observe the emergence of spontaneous condensation by impurity in other model which will be reported soon.}. In the present work, we are devoted to analysis of the low temperature region to study low-lying excitations around a ground state.

\begin{figure}[ht!] 
\begin{centering}
    \subfigure[ ]
    {\includegraphics[width=7.9cm]{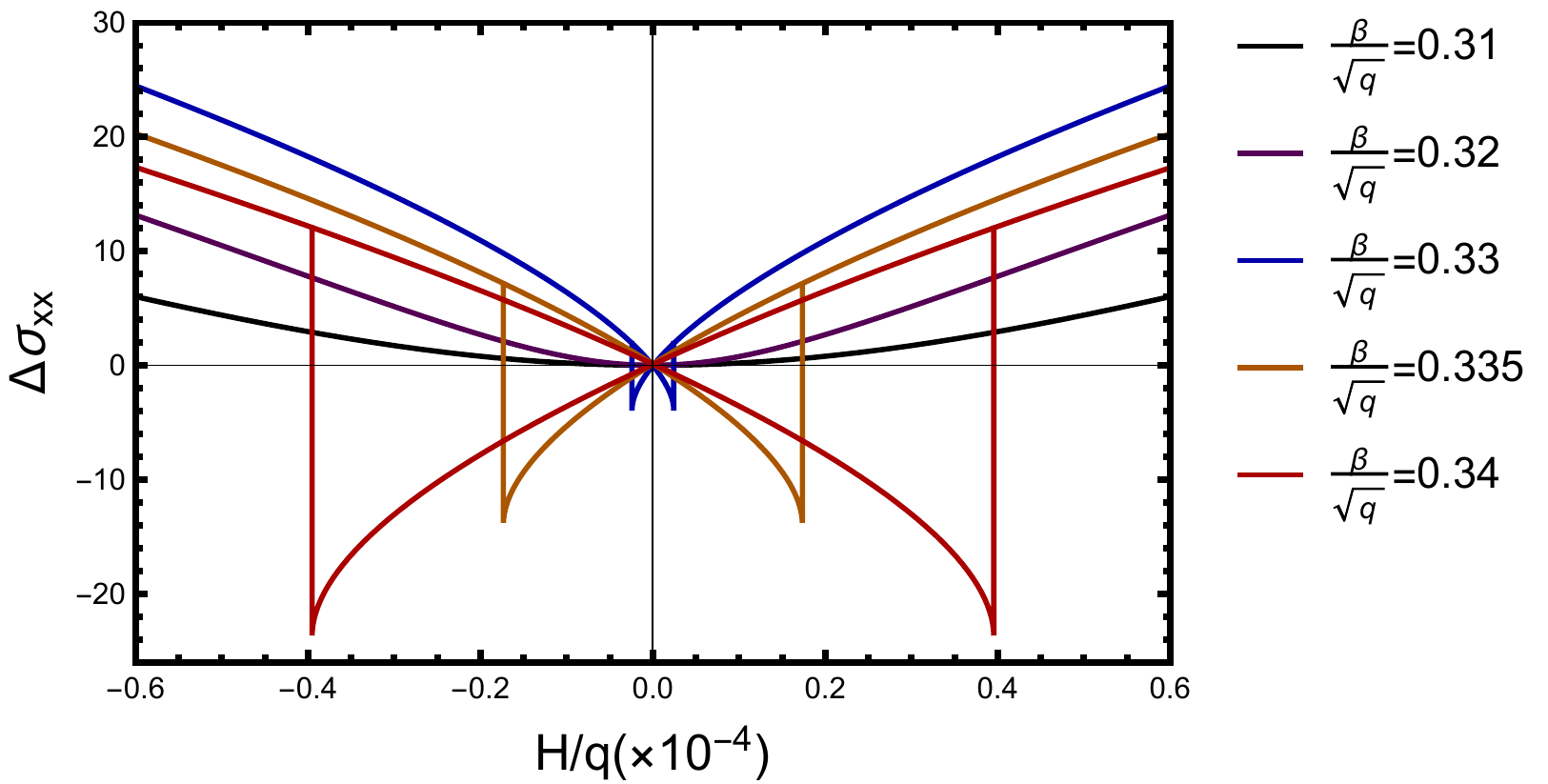}  }
       \subfigure[ ]
    {\includegraphics[width=8cm]{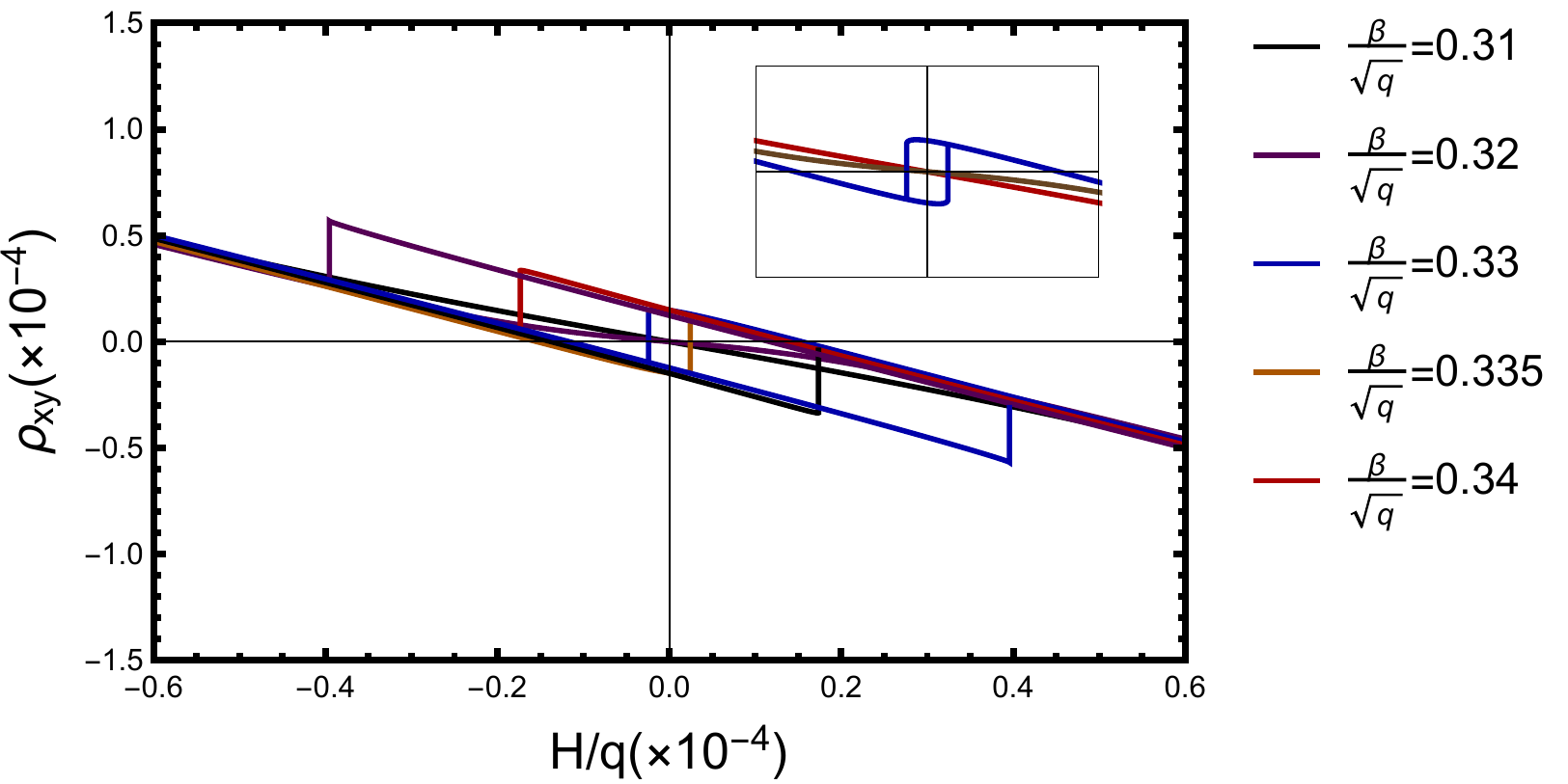}  }

    \caption{The figures show the longitudinal conductivity and Hall resistivity at $T/\sqrt{q}=3\times 10^{-4}$. These data correspond to the brown dots in Figure \ref{fig:Pdiagram}. In (a), we subtract the reference value of the conductivity so the presented value is defined by $\Delta \sigma_{xx}=\sigma_{xx} - \left(\sigma_{xx}\right)_{\mathcal{H}=0} $. These are very similar to experimental data in figure 3 of  \cite{TIexps-2}.
} \label{fig:beta_Hysteresis}
\end{centering}
\end{figure}

\begin{figure}[] 
\begin{centering}
    \subfigure[ ]
    {\includegraphics[width=7cm]{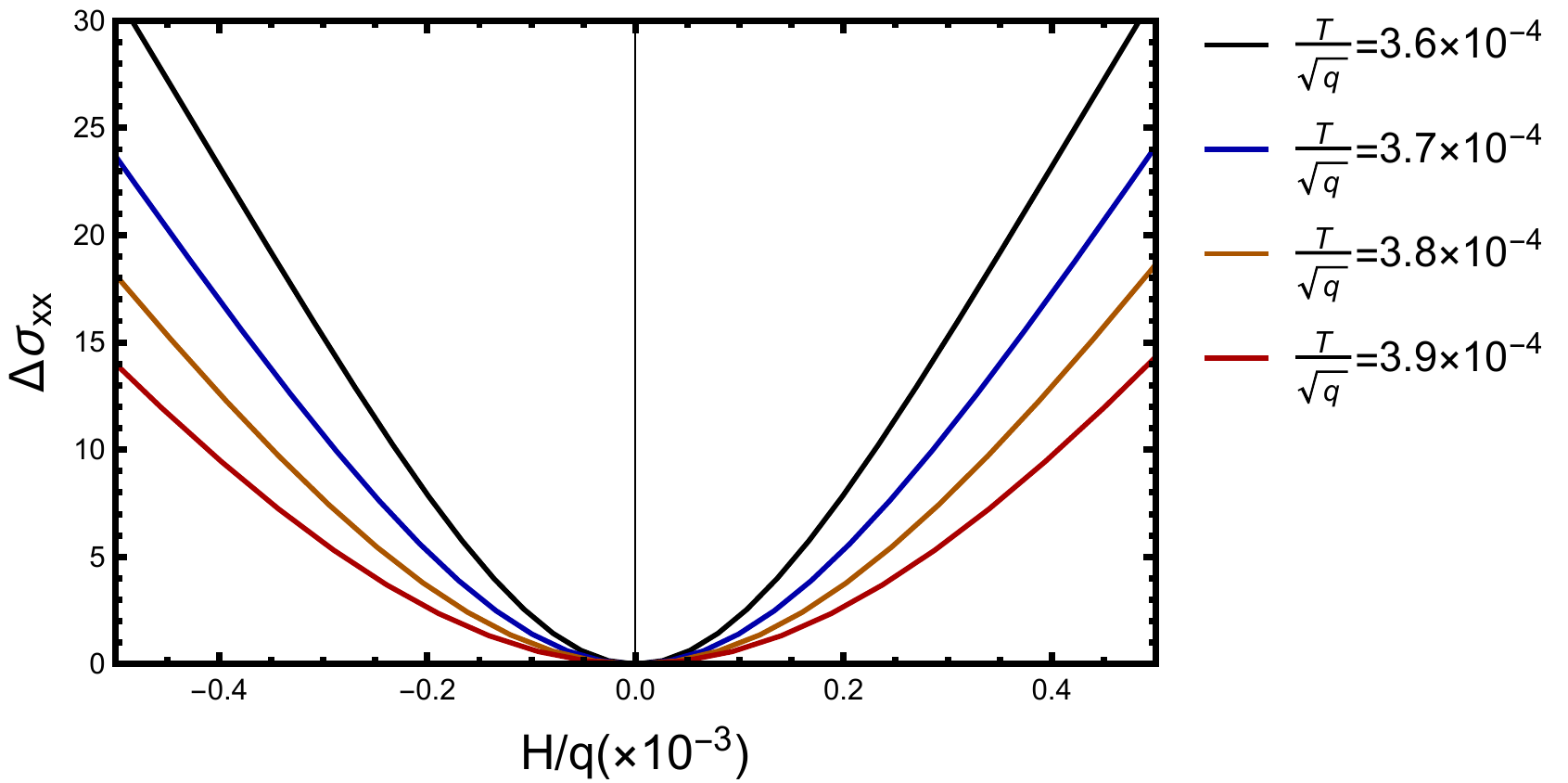}  }
    \subfigure[ ]
    {\includegraphics[width=7.2cm]{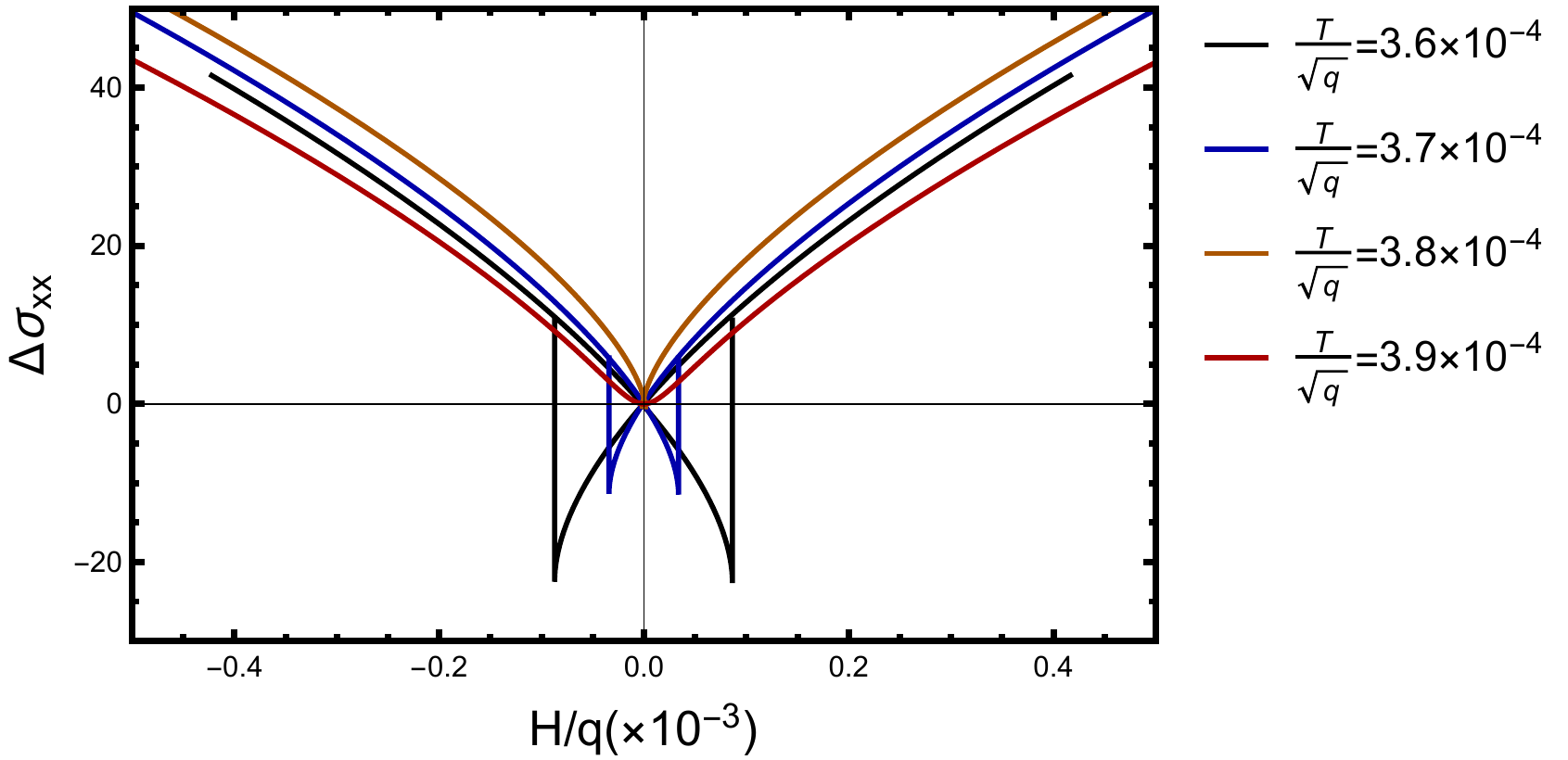}  }\\
    \subfigure[ ]
    {\includegraphics[width=7cm]{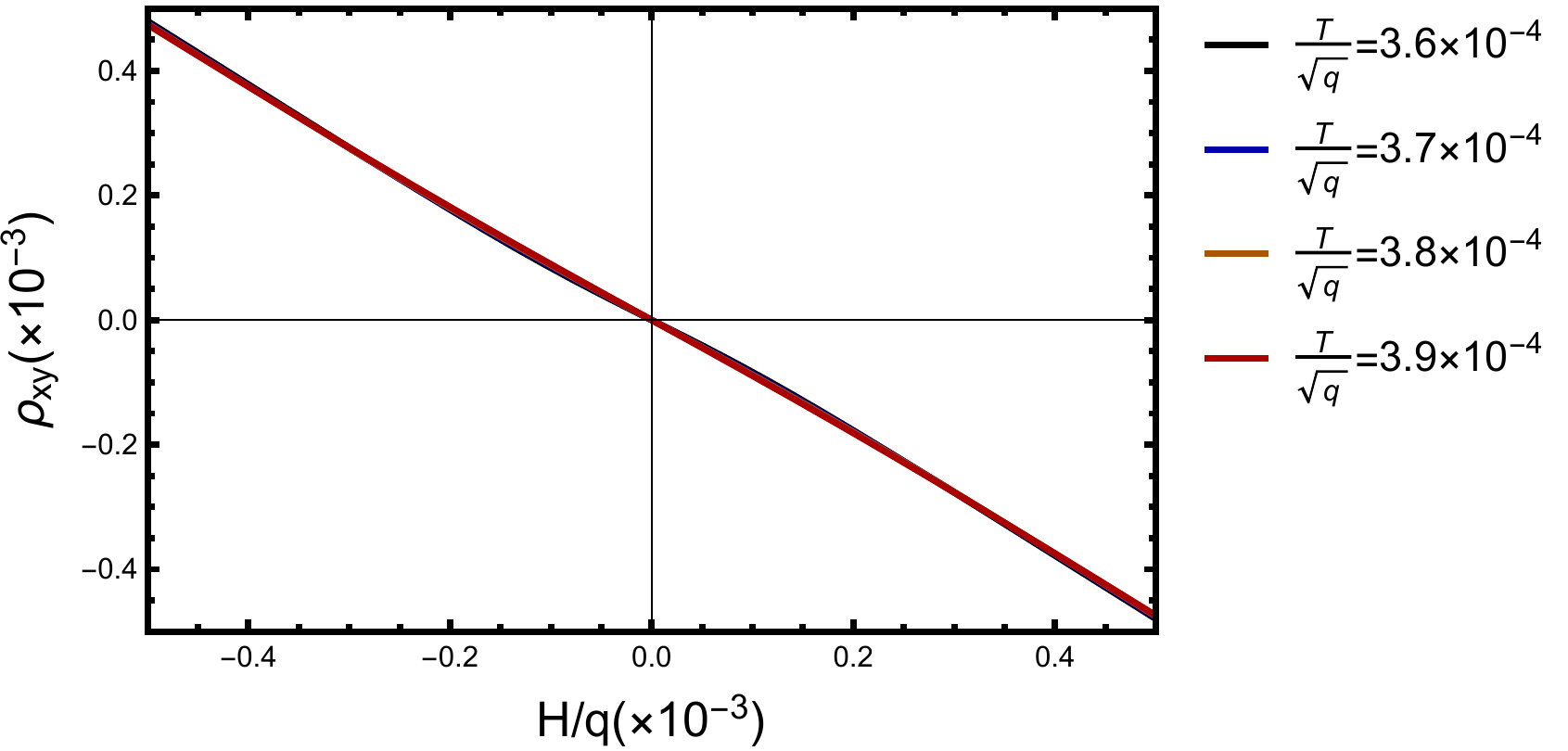}  }
    \subfigure[ ]
    {\includegraphics[width=7cm]{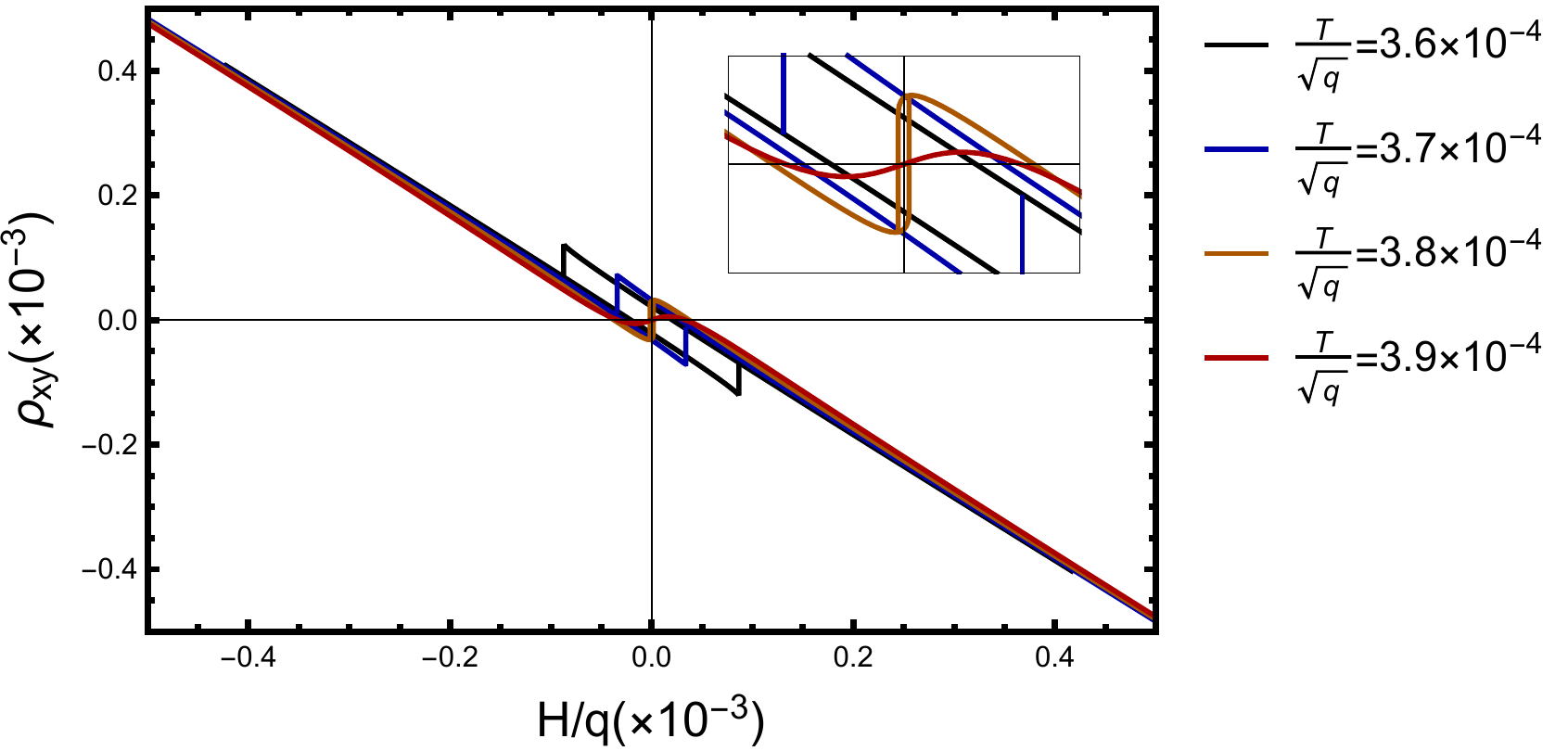}  }

    \caption{(a) and (c) corresponds to blue dots in Figure \ref{fig:Pdiagram} while (b) and (d) corresponds to red dots in Figure \ref{fig:Pdiagram}. $\Delta\sigma_{xx}$ is the subtracted conductivity like Figure \ref{fig:beta_Hysteresis}. These are also comparable to experimental data in figure 3 of  \cite{TIexps-2}.
} \label{fig:T:conductivity}  
\end{centering}
\end{figure}

We plot the hysteresis curves of the magnetoconductance in Figure \ref{fig:beta_Hysteresis} and \ref{fig:T:conductivity} under varying the impurity density and the temperature. These figures show occurrence of butterfly and loop shapes in longitudinal conductivity and Hall resistivity as $\beta$ increases or the temperature decreases. These hysteric shapes of magnetoconductance can be found in various experimental data. It is desirable to compare our result to magnetoconductance data in \cite{TIexps-1,TIexps-2}.  It seems that our holographic calculation realizes the  phase transition between WL phase and HMC phase qualitatively. We speculate that the boundary system of the model corresponds to the single magnetic domain of the material. Thus, in order to compare real experimental data quantitatively, the hysteric magnetoconductance should be averaged over all the fragments. Such a research including fragmentation average could be an interesting study but it is beyond the scope of the present work.

\begin{figure}[ht!] 
\begin{centering}
    \subfigure[ ]
    {\includegraphics[width=7cm]{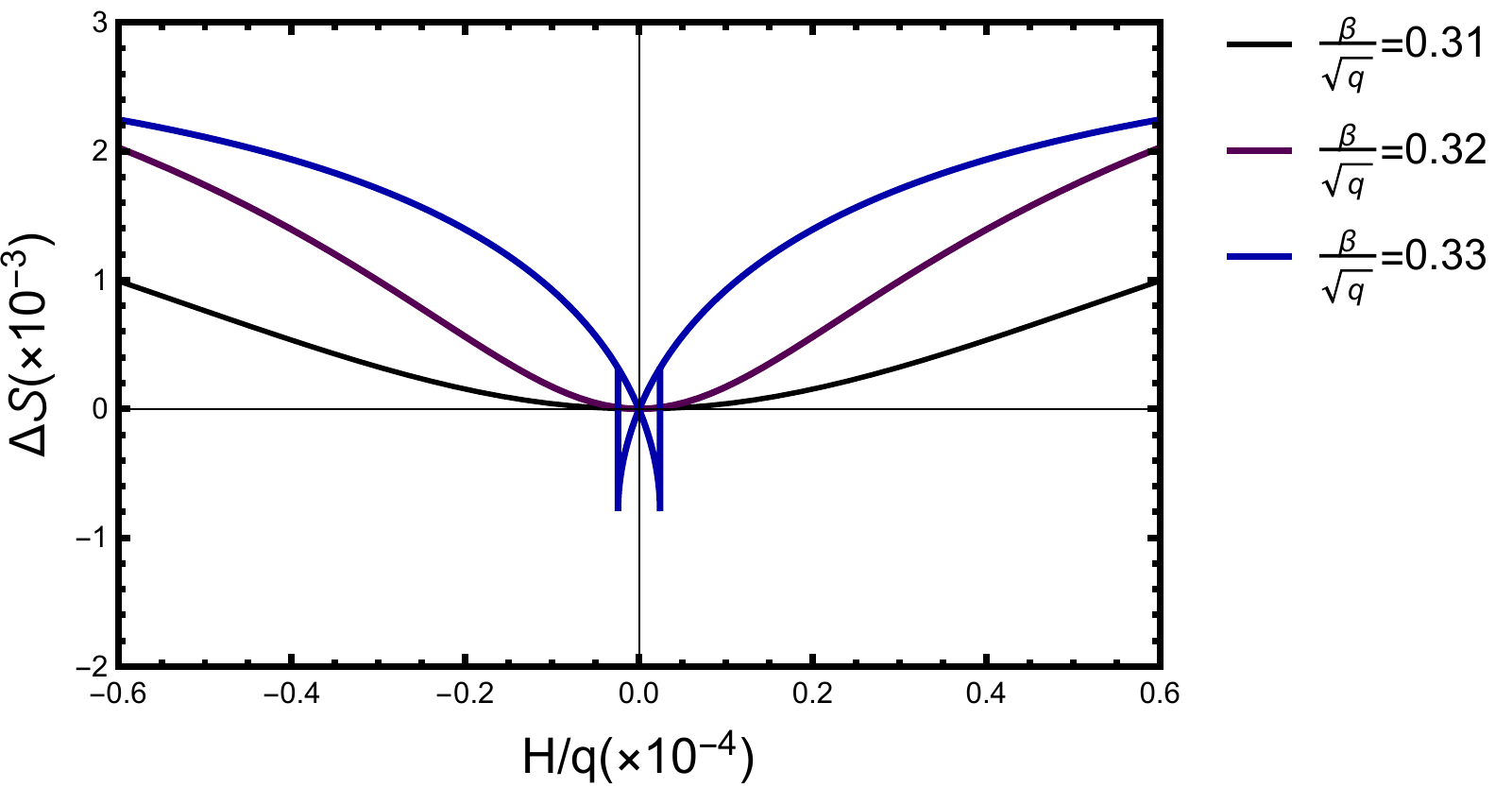}  }
    \subfigure[ ]
    {\includegraphics[width=7cm]{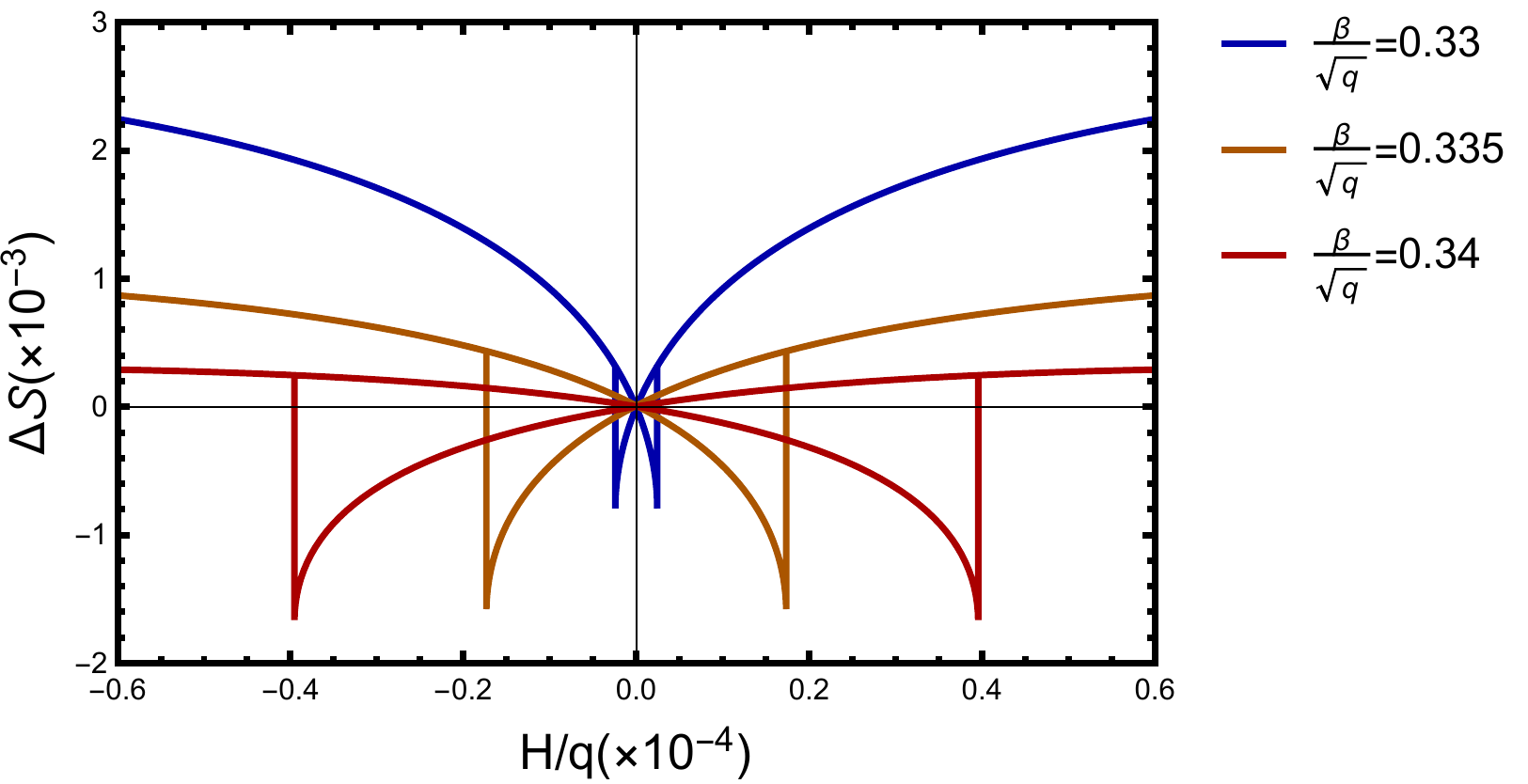}  }\\
    \subfigure[ ]
    {\includegraphics[width=7cm]{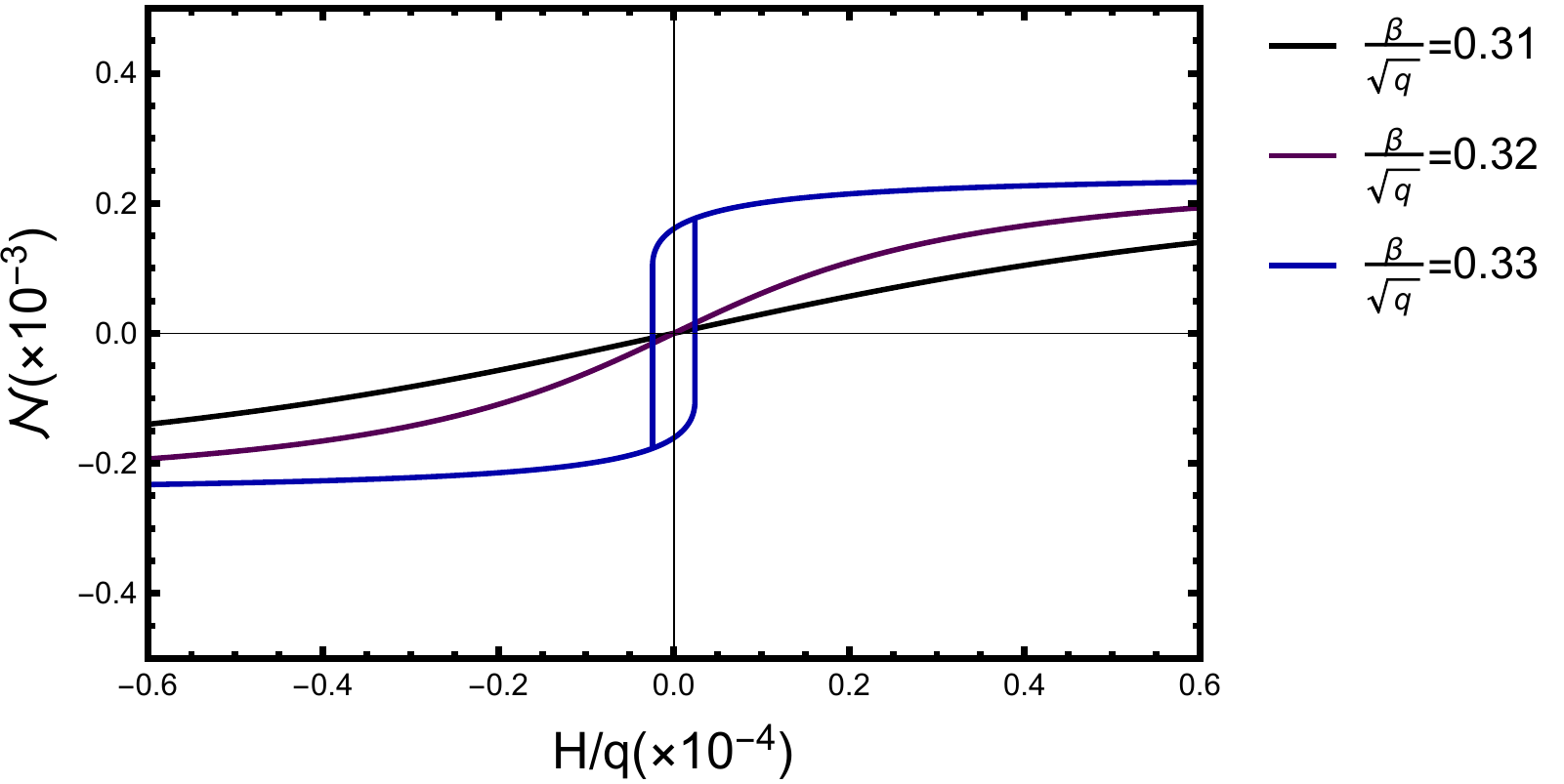}  }
    \subfigure[ ]
    {\includegraphics[width=7cm]{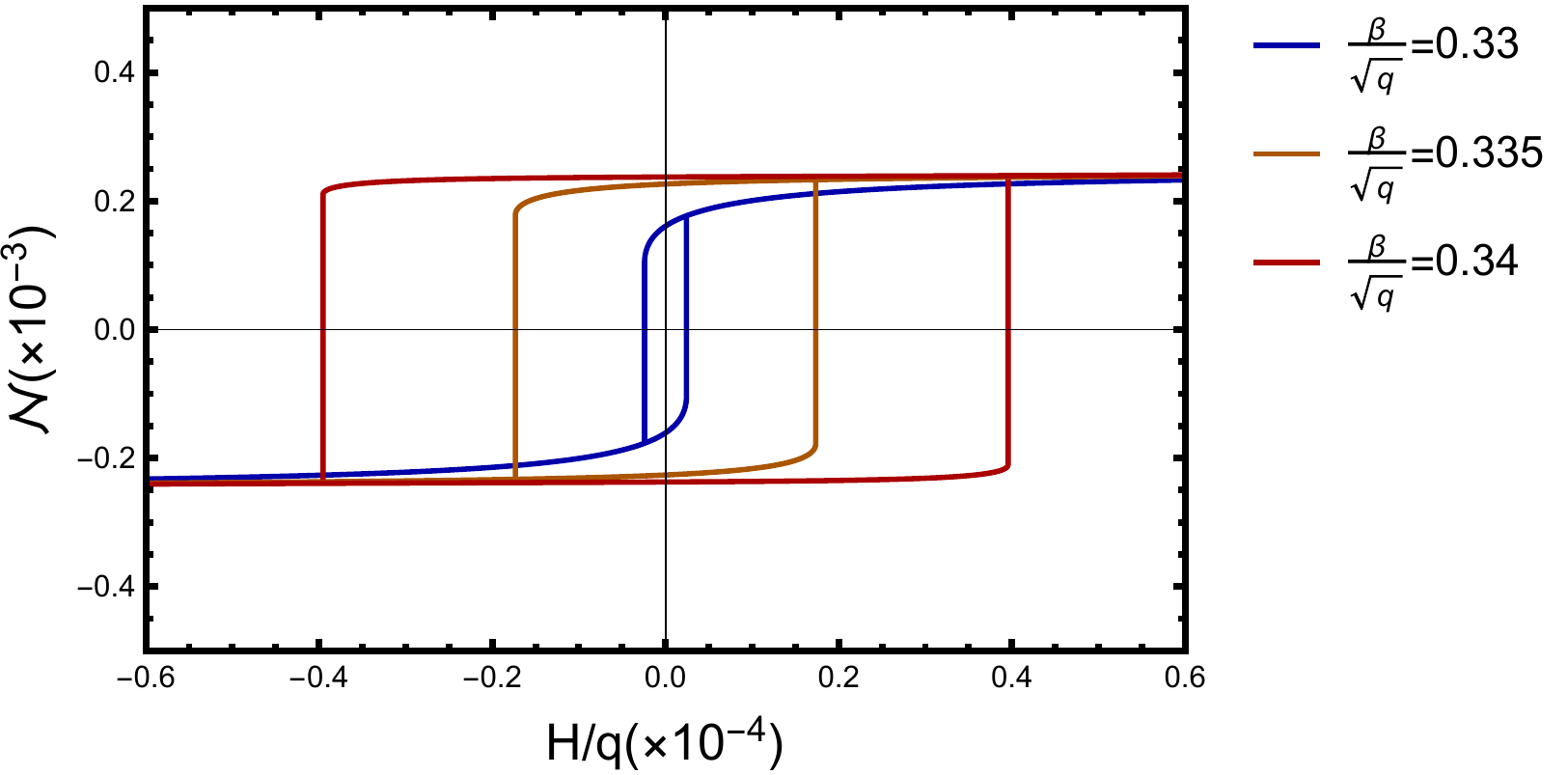}  }

    \caption{This figures show the Seebeck and the Nernst coefficients with  $T/\sqrt{q}=3\times 10^{-4}$ for the brown dots in Figure \ref{fig:Pdiagram}. Here we subtracted the reference value for the Seebeck coefficient ($\Delta \mathcal{S}\equiv\mathcal{S} - \left(\mathcal{S}\right)_{\mathcal{H}=0} $). These are our novel predictions which can be checked by experiments.
} \label{fig:beta_SN}
\end{centering}
\end{figure}

\begin{figure}[] 
\begin{centering}
    \subfigure[ ]
    {\includegraphics[width=7cm]{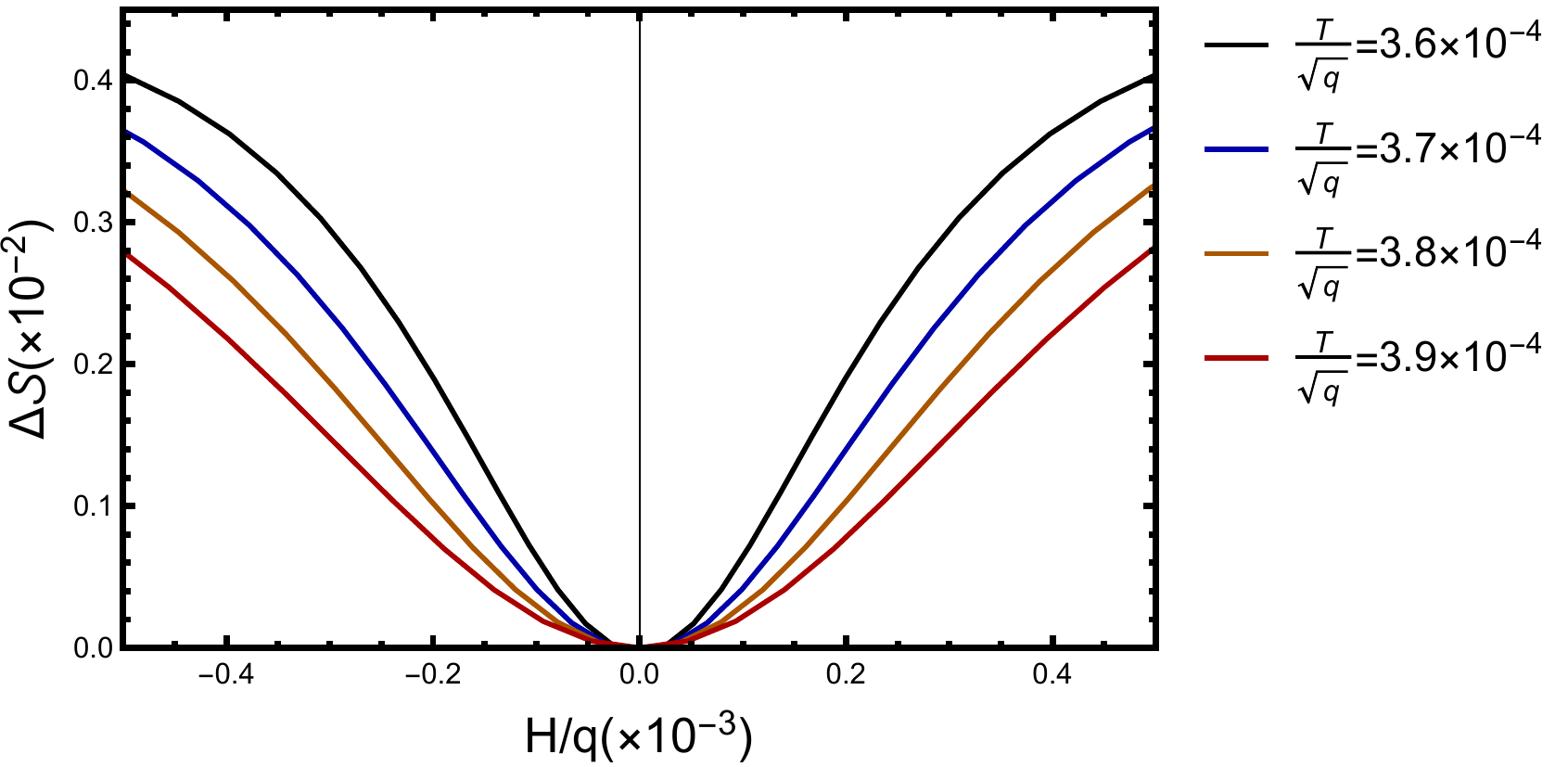}  }
    \subfigure[ ]
    {\includegraphics[width=7cm]{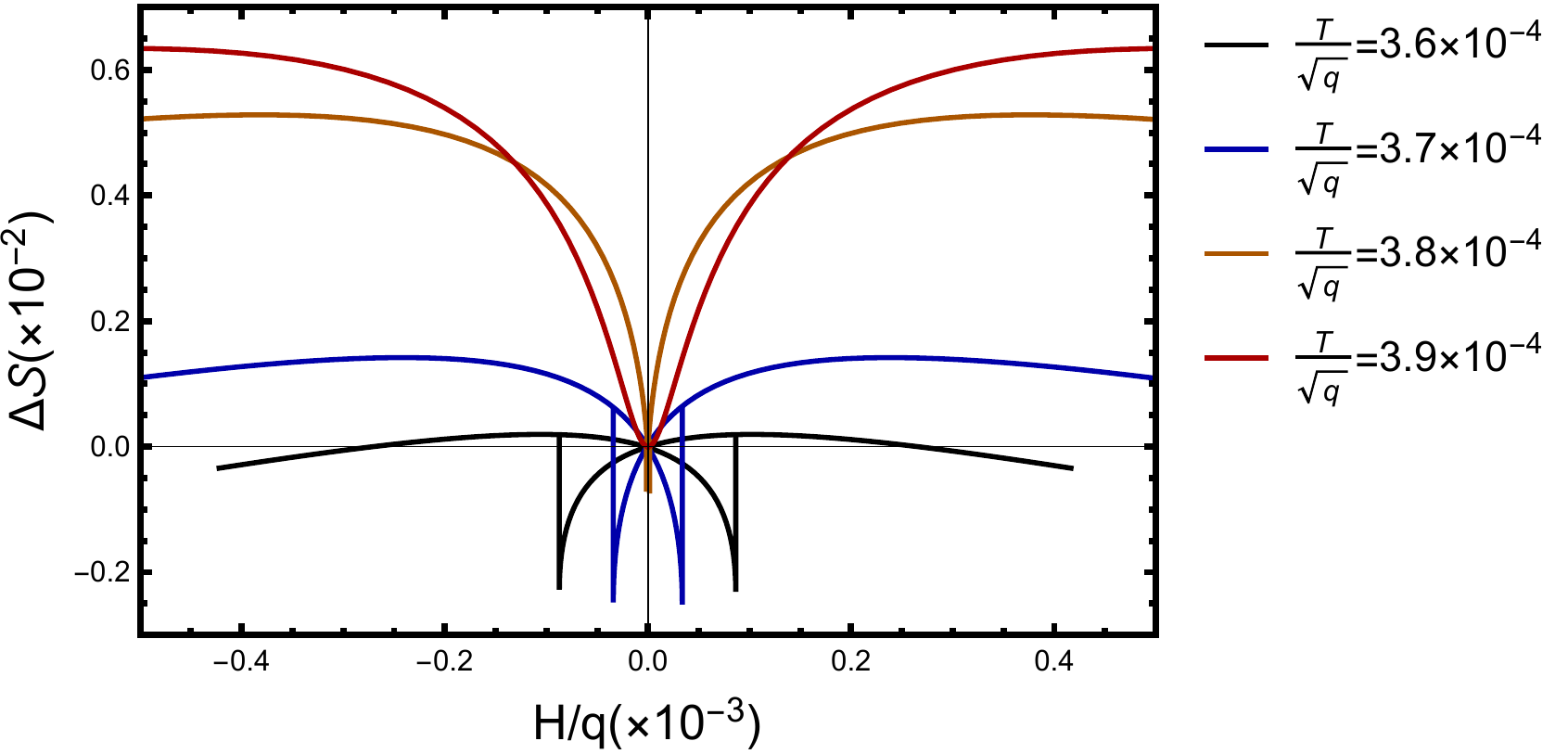}  }\\
    \subfigure[ ]
    {\includegraphics[width=7cm]{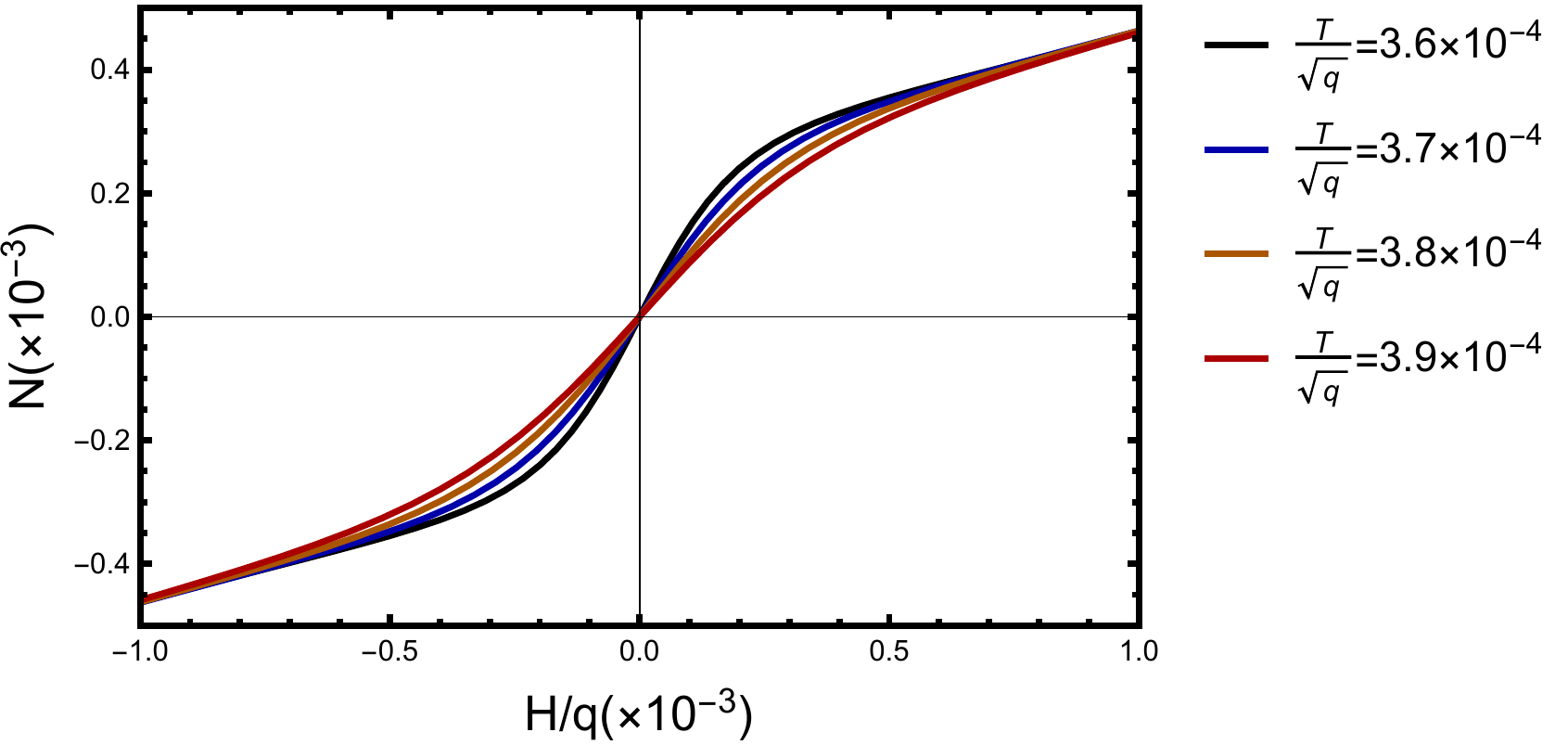}  }
    \subfigure[ ]
    {\includegraphics[width=7cm]{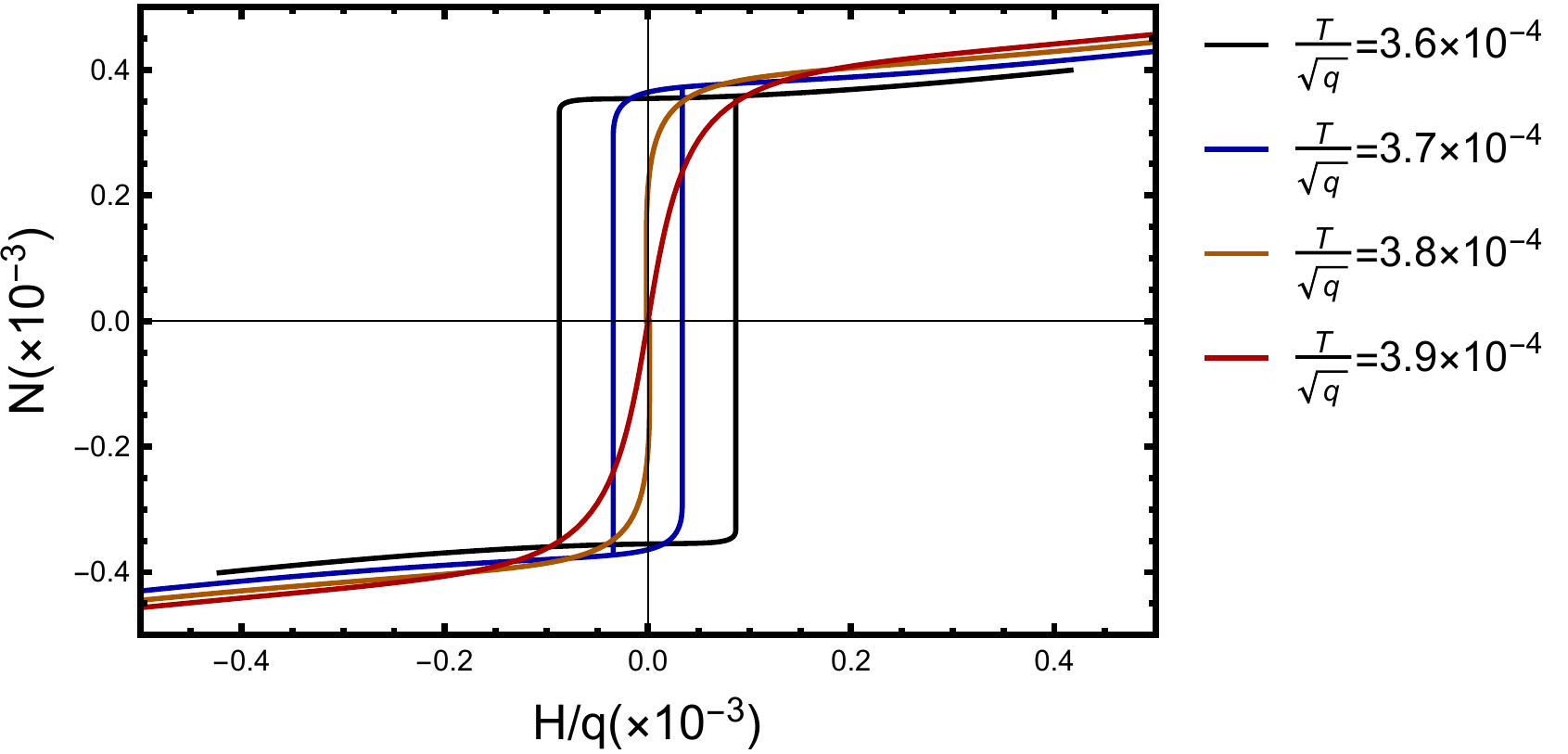}  }

    \caption{(a) and (c) corresponds to blue dots in Figure \ref{fig:Pdiagram} while (b) and (d) corresponds to red dots in Figure \ref{fig:Pdiagram}. These are our novel predictions which can be checked by experiments.
}\label{fig:T:SN} 
\end{centering}
\end{figure}

In addition to these electric conductivities, we also study hysteresis curves of the Seebeck coefficient $\mathcal{S}$ and the Nernst signal $\mathcal{N}$ which have never been measured yet in the TIs, to our knowledge. These quantities are also important to see the magnetic properties of materials, such as, anomalous Hall effect.
The Seebeck coefficient and the Nernst signal can be written in terms of the transport coefficient as follows: 
\begin{align}
\mathcal{S}\equiv \left(\rho\cdot\bar{\alpha}\right)_{xx}~,~\mathcal{N}\equiv \left(\rho\cdot\bar{\alpha}\right)_{xy}~,
\end{align} 
where $\rho_{ij}$ is the resistivity and $\bar{\alpha}_{ij}$ is the thermoelectric coefficient\footnote{$i$ and $j$ run over $x$ and $y$.}.
The appearance of the hysteric behavior of these quantities is shown in Figure \ref{fig:beta_SN} and \ref{fig:T:SN}. As one can see in the Figures, $\mathcal{S}$ and $\mathcal{N}$ show butterfly curves and hysteresis loops in the HMC phase, respectively. It would be interesting to see if such hysteric Seebeck and Nernst coefficients appear  in the same parameter regions of \cite{TIexps-1,TIexps-2}. Our derivation of the holographic conductivities is the first study to describe hysteric conductivities. The result is summarised in (\ref{conductivity}) and (\ref{alpha}). This is one of the main results in the present work.

The order parameters of this phase transition are definitely the areas of butterfly shape and hysteresis loop of the transport coefficients in the phenomenological point of view. The physical meaning of these areas is not so clear, however one may expect that these areas are related to the magnetic work given by the area of the magnetization hysteresis loop. 
 Also, it can be conjectured that these phenomenological order parameters may be related to real scalar hair and magnetization of the black brane in the holographic point of view. It can be an intersting study about the relation between the bulk and boundary quantities.

This paper is organized as follows. In section 2, we introduce a simplest holographic model to describe the magnetic hysteresis and the impurity. In section 3, we derive DC transport coefficients $\sigma_{ij}$ and $\bar{\alpha}_{ij}$ in the presence of magnetic field, axion field and real scalar hair. In section 4, we discuss our result and future directions. In Appendix
, we obtain the magnetization using the scaling symmetry technique. This independent check of the magnetization confirms that the magnetization current (\ref{Heat Current}) is the correct expression.

\section{Hysteric Black Brane with Real Scalar Hair}

As we discussed in the introduction, magnetic phenomena associated with the surface state of TIs are conjectured as strong correlation effects between the lattice and the magnetic order. Therefore, we consider gauge/gravity correspondence as a main tool to study the surface state. More explicitly, we find a dual gravity model describing the non-hysteric phase and the HMC phase on the surface of TIs. Since the gapless surface state implies the relativistic symmetry at a finite temperature, one may consider a black brane whose boundary has $SO(2,1)$ isometry. 

On the other hand, the boundary system is a 2+1 dimensional system with finite charge density together with certain impurities. The gravity dual of this system is an Einstein-Maxwell-Axion model in 3+1 dimension. Previous studies have shown that this model fits experimental data of 2-dimensional Dirac materials very well in certain range of parameter \cite{Seo:2016vks}. In addition, we introduce a neutral scalar field to the system which couples to the gauge field through a Chern-Simons like interaction which breaks TRS when scalar field is condensed. We speculate the dual boundary system is a 2+1 dimensional Dirac material with TRS breaking. Therefore, one can naturally expect that this gravity model can describe the surface state of topological insulator with impurities related to magnetic properties of the dual material.

In order to identify the physical quantities in both the gravity model and the corresponding 2+1 dimensional system, we use the standard holographic renormalization \cite{Henningson:1998gx,deHaro:2000vlm,Skenderis:2002wp}. The model admits an analytic solution for the non-hysteric phases. This solution is nothing but the dyonic black brane with an linear-axion field which is linear in the boundary spatial coordinates. On the other hand, a set of numerical solutions can be considered as the HMC phase. These numerical solutions are dyonic black branes with the linear-axion field and a real scalar hair. A set of these solutions forms a magnetization hysteresis curve. These curves without the linear-axion field were studied in \cite{Kim:2019lxb}.

A hairy black brane solution appears below a certain critical temperature. It carries nonvanishing magnetization even in the absence of the magnetic field. This nonvanishing magnetization is accompanied by vacuum expectation value of a real scalar operator as a consequence of the $\mathbb{Z}_2$ symmetry breaking in the dual field theory. Below the critical temperature, the free energy of the system has two global minima corresponding to two hairy black branes with positive and negative magnetizations, respectively. Once the system takes a global minimum spontaneously, the system follows a local minimum by turning on and varying the magnetic field. As a result, hysteresis curves of magnetization can be generated due to slowly-varying magnetic field. See \cite{Kim:2019lxb} for more details for this phenomenon.

A dual system under consideration has two spatial dimensions and a U(1) current with an external magnetic field. So we consider a holographic model with a bulk Maxwell field in asymptotically $AdS_4$ spacetime. Also, since the system undergoes the spontaneous magnetization, we include a real scalar field which can break $\mathbb{Z}_2$ symmetry of the magnetization \cite{Kim:2019lxb}. In addition to these, the momentum relaxation should come to the system in order to have finite DC conductivities. Therefore we start with the following gravity action:
\begin{align}\label{S_B}
S_B =& \frac{1}{ 16\pi G} \int d^4 x \sqrt{-g} \left( \mathcal{R}+\frac{6}{L^2} -\frac{1}{4} F^2 -\frac{1}{2}(\partial\phi)^2 - \frac{1}{2}m_{\phi}^2 \,\phi^2 - \frac{1}{2}\sum_{\mathcal{I}=1}^2 (\partial \psi^\mathcal{I})^2 \right)\nonumber\\
 & -\frac{1}{16\pi G} \int d^4 x \, \frac{\eta}{4 } \,\phi\,  \epsilon^{MNPQ} F_{MN}F_{PQ}~,   
\end{align}
where $L$ is the AdS radius and $\psi^\mathcal{I}$ describes the momentum relaxation. In addition $\eta$ represents the strength of the interaction among the real scalar, the electric field and the magnetic field. In the numerical calculation, we choose $\eta=-0.1$ for convenience.

In order to find black brane solutions admitted by this action, we take the following ansatz:
\begin{align}\label{Ansatz00}
&ds^2 = - \frac{U(r)}{L^2} e^{2W(r)-2W(\infty) } d{t}^2 + \frac{r^2}{L^2}\left( dx^2 + dy^2\right) +\frac{L^2 dr^2}{U(r)}\nonumber\\\nonumber
&\psi^\mathcal{I} =  \left(\beta\, x, \beta\, y\right)~,~\phi =\phi(r)\\
&A = A_t(r)    d{t} +\frac{\mathcal{H}}{2}L \left( x\, dy -y\, dx\right)~~.
\end{align}
By solving the Maxwell equation, one can obtain
\begin{align}\label{dAt}
A_t'(r) = e^{W(r)-W(\infty)} \frac{L^3}{r^2} \left( q   -2 \,\eta\, \phi(r)\, \mathcal{H} \right)~,
\end{align}
where $q$ is an integration constant representing the charge density of black brane. When $\phi=0$, one can find an analytic solution given by $U(r)=r^2 - \frac{\beta^2}{2}-\frac{M}{r}+\frac{\left(q^2 +\mathcal{H}^2\right)}{4 r^2} $ and $W(r)=0$, where we take a unit with $16\pi G= L =1$. This solution describes the non-hysteric phase.

Although it is difficult to find an analytic solution with nonvanishing $\phi(r)$, it is  possible to obtain numerical solutions. To do this numerical task, one may take dimensionless parameters and functions as follows:
\begin{align}
&\tilde r = \frac{r}{r_h}~~,~~U(r) = r_h^2 u(\tilde r)~,~\tilde \beta =\frac{L^2}{r_h} \beta ~,~m_{\phi}^2 = - \frac{2}{L^2}\nonumber\\
&w(\tilde r)=W(r)~,~\varphi(\tilde r) = \phi(r)~,~\tilde q =\frac{L^4}{r_h^2} q   ~,~\tilde H = \frac{L^4}{r_h^2} \mathcal{H}~~,  
\end{align}
where we took the mass square of the scalar field as $-{2}/{L^2}$ for convenience. One may try other values of the mass. So the real scalar field $\phi$ can be dual to a dimension two operator($\Delta=2$) $\mathcal{O}$. Then the equations of motion are given by:
\begin{align}
u'&+u \left(\frac{1}{4} \tilde{r} \left(\varphi '\right)^2+\frac{1}{\tilde{r}}\right)-\frac{\eta  \varphi  \tilde{H} \tilde{q}}{\tilde{r}^3}+\frac{\tilde{H}^2 \left(\eta ^2 \varphi ^2+\frac{1}{4}\right)}{\tilde{r}^3}+\frac{\tilde{q}^2}{4 \tilde{r}^3}+\frac{\tilde{\beta }^2}{2 \tilde{r}}-\left(\frac{\varphi ^2}{2}+3\right) \tilde{r}=0
\\
w'&-\frac{1}{4} \tilde{r} \left(\varphi '\right)^2=0\\
\varphi ''&-\frac{\varphi ' \left(-4 \eta  \varphi  \tilde{H} \tilde{q}+\varphi ^2 \left(4 \eta ^2 \tilde{H}^2-2 \tilde{r}^4\right)+\tilde{H}^2+\tilde{q}^2+2 \tilde{\beta }^2 \tilde{r}^2-4 u \tilde{r}^2-12 \tilde{r}^4\right)}{4 u \tilde{r}^3}\nonumber\\&+
\frac{2 \eta  \tilde{H} \tilde{q}}{u \tilde{r}^4}-\frac{\varphi  \left(4 \eta ^2 \tilde{H}^2-2 \tilde{r}^4\right)}{u \tilde{r}^4}=0~.
\end{align}
One can see that any dimensionful quantity doesn't show up in the equations of motion. By solving the above equations of motion near the boundary of AdS space, the asymptotic behavior of fields can be written as follows:
\begin{align}\label{asymptotics}
u(\tilde r) \sim\,& \tilde{r}^2+\frac{1}{4}\left(\tilde{J}^2-2 \tilde{\beta }^2
\right) -\frac{\tilde{M}}{\tilde{r}}
+\frac{2 \left(\tilde{H}^2+2 \tilde{\mathcal{O}}^2+\tilde{q}^2\right)-\tilde{J}^2 \left(\tilde{\beta }^2+\tilde{\gamma }^2\right)+\tilde{J}^4}{8 \tilde{r}^2}+\cdots\nonumber\\
w(\tilde r)\sim\,&w(\infty )-\frac{\tilde{J}^2}{8 \tilde{r}^2}+\frac{\tilde{J} \tilde{\mathcal{O}}}{3 \tilde{r}^3}+\cdots\nonumber\\
\varphi(\tilde r) \sim\,&\frac{\tilde{J}}{\tilde{r}}-\frac{\tilde{\mathcal{O}}}{\tilde{r}^2}+\frac{\tilde{J}^3}{8 \tilde{r}^3}+\cdots
\end{align}

On the other hand, we impose the following regularity condition to avoid singular configurations at the horizon:
\begin{align}
\varphi '(1) =\frac{8 \left(-2 \eta ^2 \varphi (1) \tilde{H}^2+\eta  \tilde{H} \tilde{q}+\varphi (1)\right)}{2 \left(\tilde{\beta }^2-\varphi (1)^2-6\right)+\tilde{H}^2 \left(4 \eta ^2 \varphi (1)^2+1\right)-4 \eta  \varphi (1) \tilde{H} \tilde{q}+\tilde{q}^2}~.
\end{align}
This can be achieved by an expansion of the fields near the horizon.
This condition makes the fields regular near the horizon ($\tilde{r}=1$).
Also, the temperature and entropy density can be found from mostly horizon data:
\begin{align}\label{temperature}
T =  \frac{U'(r_h)}{4\pi L^2} e^{W(r_h)-W(\infty)}=r_h \frac{u'(1)}{4\pi L^2} e^{w(1)-w(\infty)}~,~ s = \frac{ r_h^2}{4 G L^2}~.
\end{align} 
Thanks to the equations of motion, one can rewrite $u'(1)$ in terms of other parameters as follows:
\begin{align}\label{dU}
u'(1) =3-\frac{1}{4} \left(2 \tilde{\beta }^2+\tilde{H}^2+\tilde{q}^2\right)+\varphi (1)^2 \left(\frac{1}{2}-\eta ^2 \tilde{H}^2\right)+\eta  \varphi (1) \tilde{H} \tilde{q}~.
\end{align}
Since we integrated the gauge field already, $A_t$ doesn't appear explicitly. Together with $A_t(r_h)=0$, one can obtain the chemical potential by using (\ref{dAt}):
\begin{align}\label{chemi}
\mu =&  \frac{1}{L} A_t(\infty)=\frac{r_h}{L^2}\tilde{\mu} =\frac{r_h}{L^2}\int_1^\infty d\tilde{r}e^{w\left(\tilde{r}\right)-w(\infty )} \left(\frac{\tilde{q}}{\tilde{r}^2}-\frac{2 \eta  \tilde{H} \varphi \left(\tilde{r}\right)}{\tilde{r}^2}\right)~.
\end{align} 

\begin{figure}[ht!] 
\begin{centering}
    \subfigure[ ]
    {\includegraphics[width=7.8cm]{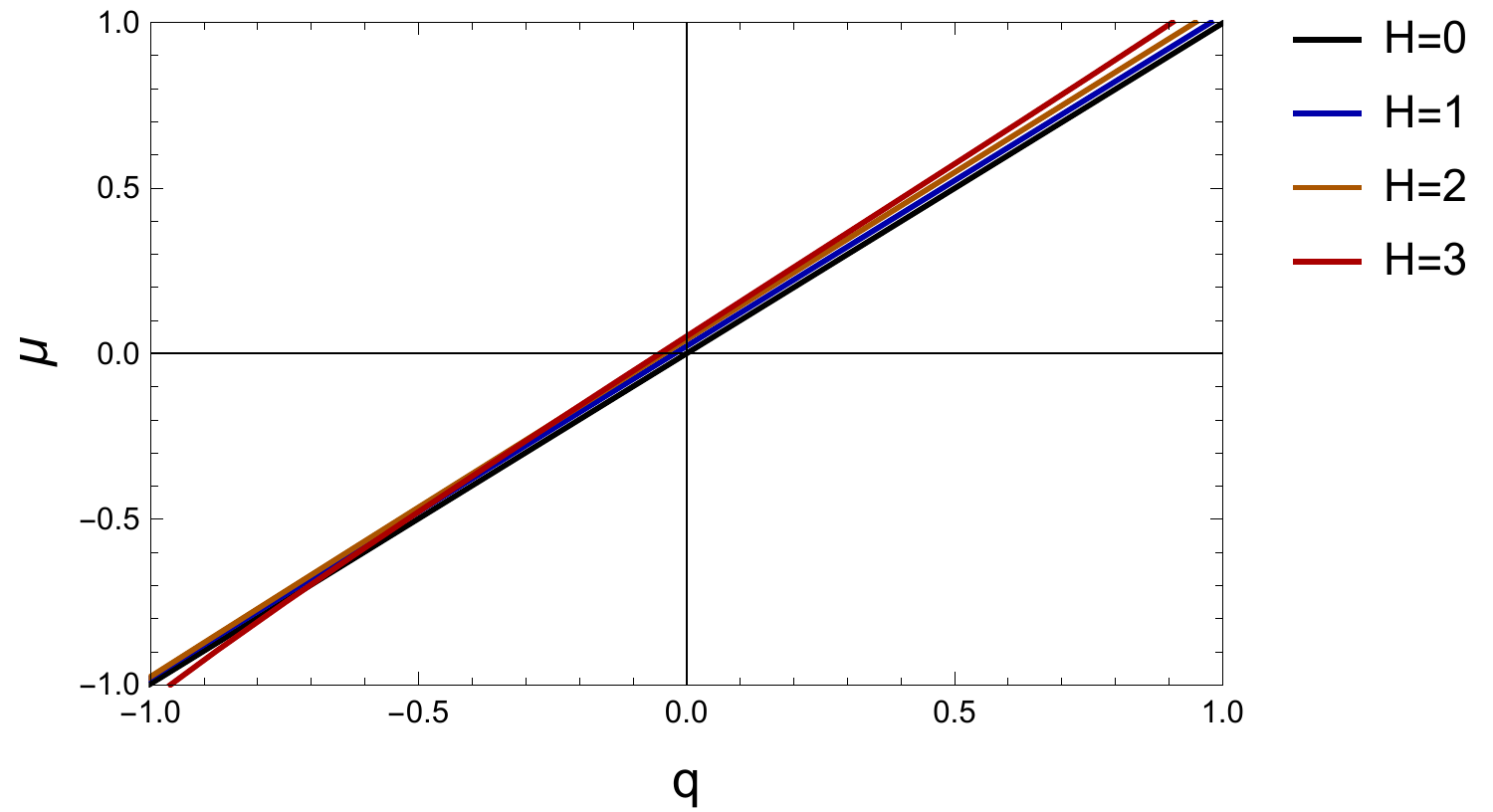}  }
       \subfigure[ ]
    {\includegraphics[width=8cm]{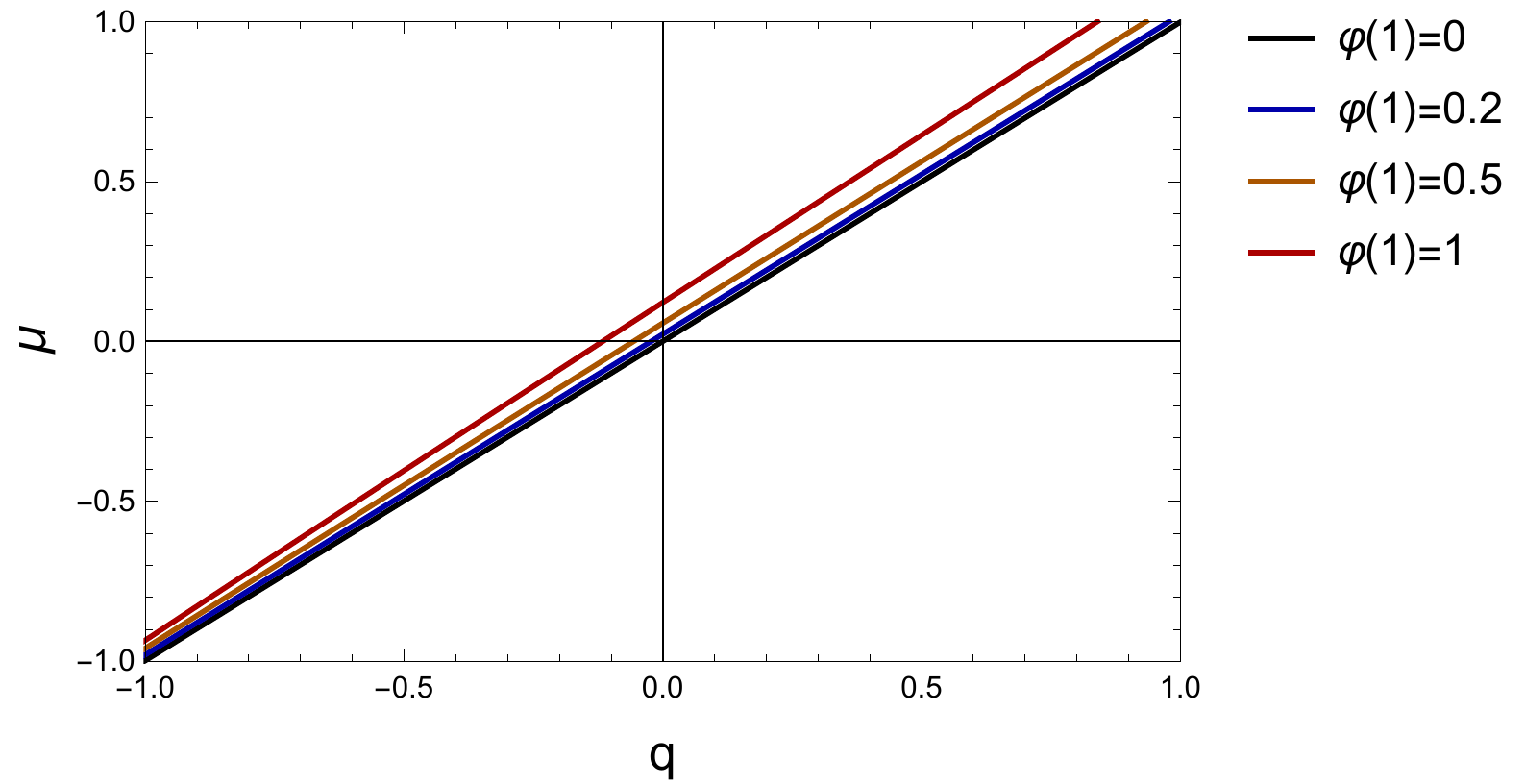}  }

    \caption{(a) Charge density dependence of chemical potential for different external magnetic field with $\varphi(1) = 0.2$ and $\beta =0.33$. (b) Charge density dependence of chemical potential for different condensation with $\beta=0.33$ and $H=1$.
} \label{fig:muq}
\end{centering}
\end{figure}

The numerical results of chemical potential (\ref{chemi}) are drawn in Figure \ref{fig:muq}. In the figure, chemical potential is almost linearly increasing in the absence of the external field or condensation according to (\ref{chemi}). When we turn on the external field or condensation, chemical potential is slightly deviated from the linear line, but the linear behavior does not change in the parameter regime of this paper. 

Following the AdS/CFT dictionary, the bulk scalar field $\phi$ corresponds to the expectation value $\left<\mathcal{O}\right>$ and the corresponding source $\mathcal J$ of an operator $\mathcal{O}$. They are identified as follows:
\begin{align}
\mathcal{J} = \frac{r_h}{L^2} \tilde{J}~,~\left<\mathcal{O}\right> = \frac{r_h^2}{16\pi L^2} \tilde{\mathcal{O}}~.
\end{align}
The powers of $r_h$ represent the dimensions of the source and the operator in the dual field theory. Since we would like to consider a simplest construction, we pick solutions with $\mathcal{J}=0$. Using the above consideration, it is possible to find solutions of the equations of motion numerically. The obtained solutions have been used to draw the phase diagram in Figure \ref{fig:Pdiagram} and to evaluate the magnetoconductance in Figure \ref{fig:beta_Hysteresis}-\ref{fig:T:SN}.

\section{Holographic Conductivity}\label{sec:HConductivity}\label{DCconductivity}

Now, we try to find DC conductivities in the dual field theory using holographic approach developed in \cite{Andrade:2013gsa,Donos:2014cya,Blake:2015ina,Kim:2015wba,Seo:2015pug,Seo:2016vks}. These holographic conductivities can be obtained by solving the equations of motion linearly responding to an external electric field, ($E_{x}$,$E_y$). Thus we consider relevant metric fluctuations as follows:
\begin{align}\label{feq1}
ds_{(1)}^{2}=2\lambda \left\{ \delta g_{tx}(r)dt dx+ \delta  g_{ty}(r) dt dy  + r^2 \delta h_{rx}(r) dr dx +   r^2 \delta h_{ry}(r)dr dy   \right\}~,
\end{align}
where $\lambda$ is a formal expansion parameter. Together with this fluctuation, the matter fields also has corrections given by 
\begin{align}\label{feq2}
A_{(1)} = \lambda \big\{ (- E_x L t + \delta A_x(r) ) dx  +(- E_y L t + \delta A_y(r) ) dy    \big\}
\end{align}
for the gauge field and
\begin{align}\label{feq3}
\psi_{(1)}^{\mathcal I} = \lambda ( \delta\psi_x(r), \delta\psi_y(r) )~
\end{align}
for the linear-axion fields $\psi^{\mathcal I}$. From the vector property of the fluctuation, one can see that there is no real scalar ($\phi$) fluctuation at the linear level.

To find the first order solution in $\lambda$, we have to impose an appropriate boundary condition at the horizon. The correct boundary conditions is nothing but the in-going boundary condition and the explicit form is as follows:
\begin{align}\label{ingoing}
&\delta g_{tx}(r) = \delta g^0_{tx} + \mathcal{O}\left(r-r_h \right)~,~\delta g_{ty}(r) = \delta g^0_{ty} + \mathcal{O}\left(r-r_h \right)~\nonumber\\
&\delta h_{tx}(r) \sim \frac{L^2}{r^2 U(r)} \delta g_{tx}^0~,~\delta h_{ty}(r) \sim \frac{L^2}{r^2 U(r)} \delta g_{ty}^0~,\nonumber\\
&\delta A_x(r) \sim - E_x \frac{L}{4\pi T} \log (r-r_h)~,~\delta A_y(r) \sim - E_y \frac{L}{4\pi T} \log (r-r_h)~,
\end{align}
where ``$\sim$" means up to the singular term near the horizon. In addition $\psi^{\mathcal I}$ can be taken as a regular field near the horizon.

On the other hand, the holographic $U(1)$ current and the heat current through gauge/gravity duality are given by\footnote{$\mu$ runs over 0,1,2.} 
\begin{align}\label{Current}
&J^\mu =\lim_{r\to\infty} \frac{L}{16\pi G}\left(\sqrt{-g}F^{\mu r} -\eta\, \phi\,\epsilon^{r\mu\nu\sigma}F_{\nu\sigma}   \right),\\
&Q^\mu= T^{t\mu}-\mu J^\mu~,
\end{align}
where $T^{t\mu}$ denotes components of the holographic energy-momentum tensor $T^{\mu\nu}$. $J^\mu$ and $T^{\mu\nu}$ are obtained by derivatives of the on-shell action with respect to $\frac{1}{L}A_\mu(\infty)$ and the boundary metric.

In fact, $J^\mu$ is independent of the radial coordinate $r$ due to the Maxwell equation. Therefore, one can find the holographic current in terms of horizon behavior of the fields. The resultant expression $J^\mu$ depends on the constants, $\delta g_{tx}^0$ and $\delta g_{ty}^0$ appearing in (\ref{ingoing}). By solving $(x,r)$ and $(y,r)$-components of the first order Einstein equation in $\lambda$ with the regularity condition (\ref{ingoing}), one can find $\delta g_{tx}^0$ and $\delta g_{ty}^0$ in terms of the other parameters. Plugging the expression of $\delta g_{tx}^0$ and $\delta g_{ty}^0$ into the holographic current $J^\mu$ in (\ref{Current}), one can obtain the electric conductivity defined by $J^i = \sigma_{ij}E_{j}$ as follows: 
\begin{align}\label{conductivity}
\sigma_{xx} =\frac{L^2}{16\pi G} \frac{\mathcal{N}_{xx}}{\mathcal{D}_{xx}}~,~\sigma_{xy} =\frac{L^2}{16\pi G} \frac{\mathcal{N}_{xy}}{\mathcal{D}_{xx}}~,~\sigma_{yx} = -\sigma_{xy}~,~\sigma_{yy}=\sigma_{xx}~,
\end{align}
where
\begin{align}
\mathcal{D}_{xx} =&e^{2w(\infty)}\left(\tilde{q}-2 \eta  \varphi (1) \tilde{H}\right)^2 \tilde{H}^2+e^{2 w(1)} \left(\tilde{H}^2+\tilde{\beta }^2\right)^2\nonumber\\
\mathcal{N}_{xx} =& e^{2 w(1)} \tilde{\beta }^2 \left(\tilde{\beta }^2+\tilde{H}^2\right) +\tilde{\beta }^2e^{w(\infty )+w(1)} \left(\tilde{q}-2 \eta  \varphi (1) \tilde{H}\right)^2\nonumber\\
\mathcal{N}_{xy} =&\tilde{\beta }^2 \tilde{H} e^{w(\infty )+w(1)} \left(\tilde{q}-2 \eta  \varphi (1) \tilde{H}\right) + \tilde{H} e^{2 w(\infty )} \left(\tilde{q}-4 \eta  \varphi (1) \tilde{H}\right) \left(\tilde{q}-2 \eta  \varphi (1) \tilde{H}\right)^2   \nonumber 
\\& -e^{2 w(1)} \left(\tilde{\beta }^2+\tilde{H}^2\right) \left(2 \eta  \varphi (1) \tilde{\beta }^2+4 \eta  \varphi (1) \tilde{H}^2-\tilde{H} \tilde{q}\right)~.
\end{align} 

On the other hand, in order to obtain the heat current $Q^i\equiv T^{ti}-\mu J^i$, we define the bulk heat current as follows:
\begin{align}\label{eq:bulk_heat}
\hat{Q}^i \equiv\frac{1}{16\pi G L^2} U(r)^2 e^{3W(r)-3W(\infty)} \delta^{ij}  \left( \frac{g_{tj}(r)}{U(r)e^{2W(r)-2W(\infty)}} \right)' -\frac{1}{L} A_t(r) J^i \,.
\end{align}
Using the equations of motion at the linear level, one can verify that
\begin{align}
(\hat{Q}^i)' =- \epsilon^{ij} \left( \bar{m}  -  (\frac{\eta L}{4\pi G} \phi A_t )' \right) E_j~,
\end{align}
where
\begin{align}
\bar{m}= \frac{L^4}{8 \pi  G}e^{W(r)-W(\infty )}\left(\frac{\eta\,  q \phi(r)}{r^2} -\frac{\mathcal{H} \left(1+4 \eta ^2 \phi(r)^2\right)}{2 r^2} \right)~.
\end{align}
Taking integration from the horizon to the boundary of AdS space, the following relation can be evaluated:
\begin{align}\label{Heat Current}
\hat{Q}^i(\infty)=\hat{Q}^i(r_h) -  \epsilon^{ij} \mathcal{M}_{\mathcal{H}} E_j~~\text{with}~\mathcal{M}_{\mathcal{H}}= \int_{r_h}^\infty dr \,\bar{m}.
\end{align}
One can also show that the heat current $Q^i$ is same with the bulk current $\hat{Q}^i$ at the boundary of the AdS space, i.e, $Q^i =\hat{Q}^i(\infty)$. Therefore, the holographic heat current $Q^i$ is given by the horizon quantities and the magnetization current. Here it is important to note that the magnetization current, given by the last term, provides the explicit form of magnetization $\mathcal{M}_{\mathcal{H}}$. Usually the magnetization can be obtained by the first law of black hole thermodynamics but we didn't use the first law. Thus, this evaluation of the heat current is another derivation of magnetization without using the first law. In Appendix
 we find the expression of the magnetization (\ref{eq:magnetization}) using the scaling symmetry technique. Such an independent check confirms that our derivation of heat current leads to the correct expression.

Therefore, the holographic heat current $Q^i$ is given by the horizon quantities and the magnetization current. However the magnetization current is the bound current which can not be measured. Thus, the transport coefficient can be read off from $\hat{Q}^i(r_h)$. The subtracted heat current can be found as follows:
\begin{align}
Q^i+\epsilon^{ij} \mathcal{M}_{\mathcal{H}} E_j = - \frac{1}{16\pi G L^2} e^{W(r_h)-W(\infty)} U'(r_h) g_{ti}(r_h)= -\frac{1}{4G} T \delta g_{ti}^0.
\end{align} 
Using the explicit expression of $\delta g_{ti}^0$ determined by the Einstein equation, we arrive at the final form of the thermo-electric coefficient\footnote{Thermoelectric conductivities are defined by 
$
J^i = \sigma_{ij}E^j~,~Q^i =\bar{\alpha}_{ij} T E^j-\epsilon^{ij}\mathcal{M}_{\mathcal{H}}E_j
$.}: 
\begin{align}\label{alpha}
\bar{\alpha}_{xx}= \frac{L^2}{4G}\frac{\mathbb{N}_{xx}}{\mathcal{D}_{xx}}~,~\bar{\alpha}_{xy}= \frac{L^2}{4G}\frac{\mathbb{N}_{xy}}{\mathcal{D}_{xx}}~,~\bar{\alpha}_{xx}=\bar{\alpha}_{yy}~,~\bar{\alpha}_{yx}=-\bar{\alpha}_{xy}~,~
\end{align}
where 
\begin{align}
&\mathbb{N}_{xx}=\left(\tilde{q}-2 \eta  \varphi (1) \tilde{H}\right) \left(e^{w(\infty )+w(1)} \left(\tilde{\beta }^2+\tilde{H}^2\right)-\tilde{H}^2 e^{2 w(\infty )}\right),\nonumber\\
&\mathbb{N}_{xy}=\tilde{H} \left(e^{2 w(\infty )} \left(\tilde{q}-2 \eta  \varphi (1) \tilde{H}\right)^2+e^{w(\infty )+w(1)} \left(\tilde{\beta }^2+\tilde{H}^2\right)\right)~.
\end{align}

In the expressions (\ref{conductivity}) and (\ref{alpha}), one can see that we need the asymptotic value of $w(\tilde{r})$. However, it is possible to find the solution satisfying $w(\infty)=0$ by a shooting method. So the magnetoconductance ($\sigma_{ij}$ and $\bar{\alpha}_{ij}$) is given by the horizon data of black brane. We plot the numerical results in Figure \ref{fig:beta_Hysteresis}-\ref{fig:T:SN} in the unit of $16\pi G =L =1$. The figures present the magnetoconductance with various parameters corresponding to dots in the phase diagram of Figure \ref{fig:Pdiagram}.

\begin{figure}[] 
\begin{centering}
    \subfigure[ ]
    {\includegraphics[width=7cm]{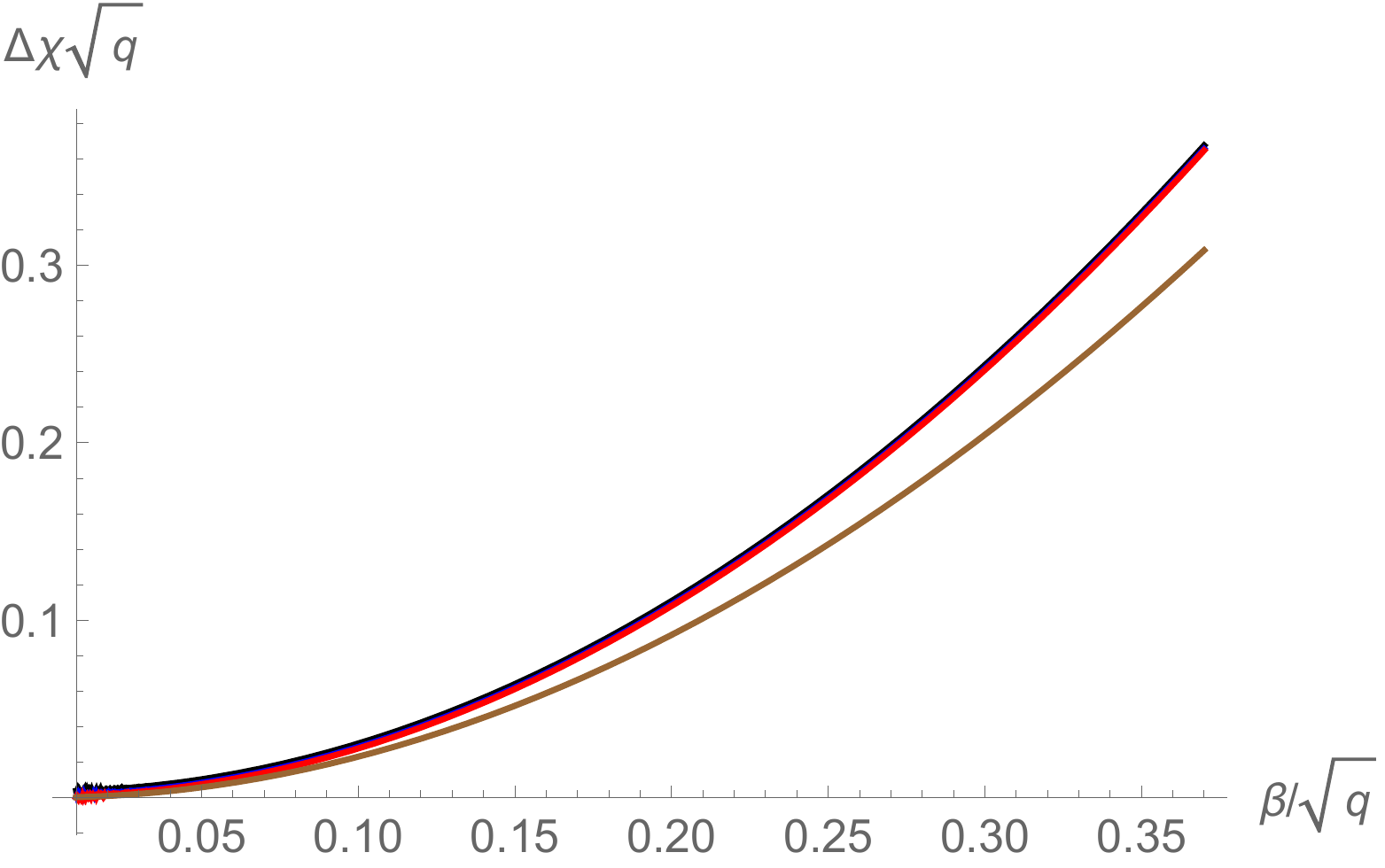}  }
        \subfigure[ ]
    {\includegraphics[width=8.8cm]{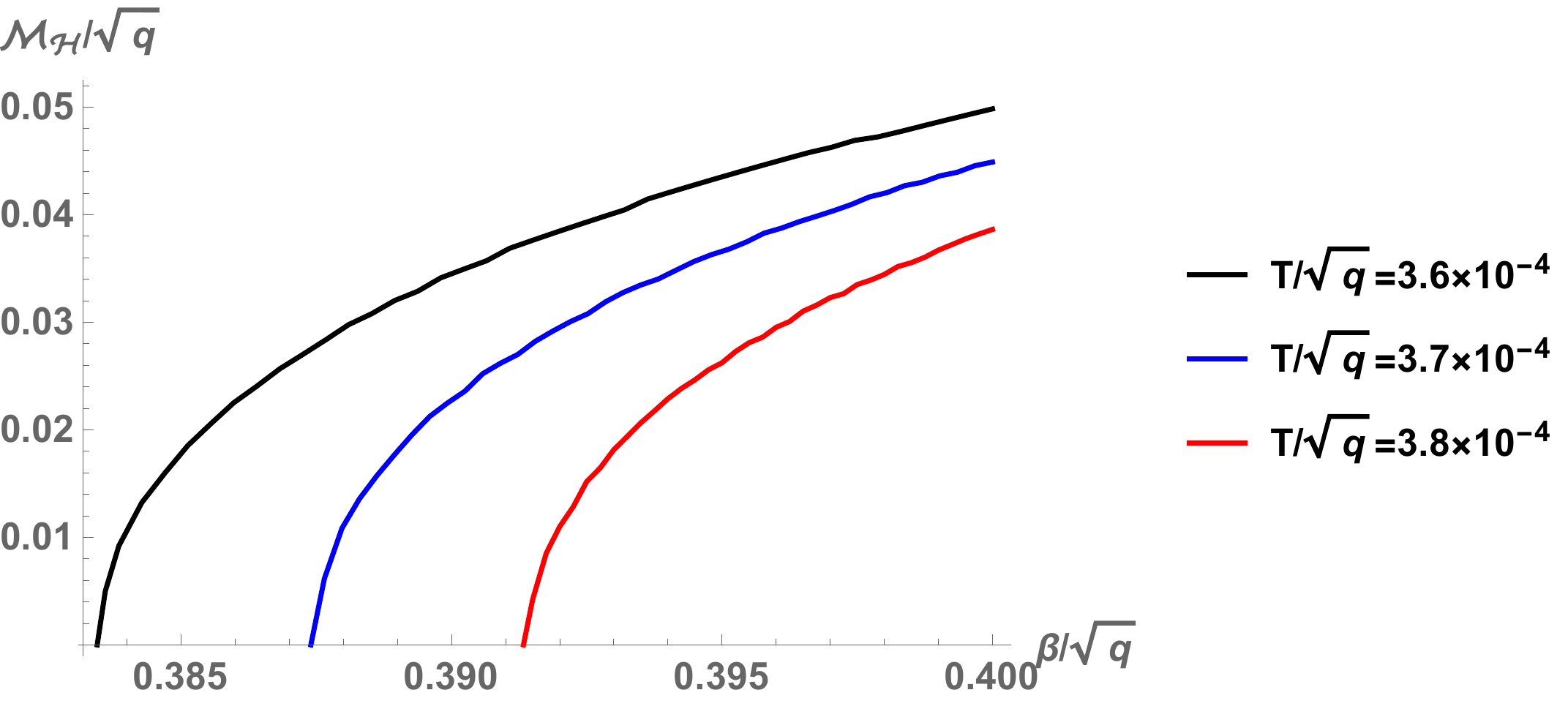}  }
       
    \caption{(a) shows the susceptibility (\ref{susceptibility}) in the WL phase. To see increasing behavior clearly, we subtract the reference value so the presented curves for $\Delta \chi\equiv \chi-\chi_{\beta=0}$. The brown curve ($T/\sqrt{q}=1.5\times 10^{-2}$) added to show the coincidence of the other curves is just from the small parameter differences. (b) shows the spontaneous magnetization in the HMC phase. The spontaneous magnetization begins to appear at certain critical impurity densities and grows with increasing impurity.
} \label{fig:MvsBeta}
\end{centering}
\end{figure}

As an important argument, we explain how the impurity density $\beta$ can be related to magnetic properties in Figure \ref{fig:MvsBeta}. The impurity density $\beta$ in (\ref{Ansatz00}) does not interact with the magnetic field directly but it can interact with the magnetic field via bulk back-reaction indirectly, as we mentioned in the introduction. In order to see such an interaction, one can take into account relevant physical quantities. Since there is no spontaneous magnetization in the non-hysteric phase, one may consider the susceptibility given by 
\begin{align}\label{susceptibility}
\chi= \frac{\partial \mathcal{M}_{\mathcal{H}}}{\partial \mathcal{H}}|_{\mathcal{H}=0} = -\frac{L^2}{16\pi G}\frac{1}{r_h}~,
\end{align}
where we took $\mathcal{H}=0$ to make comparison parallel with the spontaneous magnetization in the HMC phase. In this case, we can get the relation between the magnetic susceptibility and the other parameters using (\ref{temperature}) and (\ref{dU}) as
\begin{align}\label{eq:chi}
\frac{q^2}{4} \chi^4 +\frac{\beta^2}{2} \chi^2 -4\pi T \chi -3 =0. 
\end{align}
 Here, we can clearly see that the impurity density $\beta$ directly affects to the magnetic susceptibility. We can get analytic solution of the algebraic equation (\ref{eq:chi}) which is very complicated. We expand the solution of (\ref{eq:chi}) in small impurity density limit;
 \begin{align}
\Delta \chi |_{\beta << 1} &\sim  \left( \frac{3 \sqrt{3}-\left(\sqrt{3}+2\right) \pi ^2 T^2}{6 \sqrt{2}\, 3^{3/4}}+\cdots \right) \frac{\beta^2}{q} + {\cal O}(\frac{{\beta^4}}{q^2})~~~~{\rm for}~ T/\sqrt{q} << 1 \cr
 &\sim \left( \frac{9}{128 \pi ^3 T^3} + \cdots \right) \frac{\beta^2}{q} + {\cal O}(\frac{\beta^4}{q^2})~~~~{\rm for}~ T/\sqrt{q}>>1,
 \end{align}
 where $\Delta \chi\equiv \chi-\chi_{\beta=0}$.
 The results of full solution are shown in Figure \ref{fig:MvsBeta} (a). 

On the other hand, the spontaneous magnetization $\mathcal{M}_{\mathcal{H}}$  without other sources for a fixed charge density 
 is the most relevant quantity to see the magnetic effect from $\beta$ in the HMC phase. The numerical results are shown in Figure \ref{fig:MvsBeta} (b). In summary, the susceptibility and the magnetization increase with increasing $\beta$. Therefore, we conclude that $\beta$ is indeed related to magnetic properties. Even though it is not still clear this impurity really plays a role of magnetic impurity, this may help to construct magnetic impurities in holographic models.

\section{Conclusion}

In this paper we constructed a model which shows a novel phase transition between the non-hysteric WL phase and the HMC phase. To describe this phase transition, we derived the holographic magnetoconductance formulas (\ref{conductivity}) and (\ref{alpha}) for the hairy black brane with the linear-axion field and the real scalar hair. Using these formulas and numerical solutions, we confirm that the electric conductivity $\sigma_{ij}$ shows the phase transition reported in the earlier experiments \cite{TIexps-1,TIexps-2}. Also, we evaluated the Seebeck coefficient and the Nernst signal. These quantities also show the hysteric behavior in the HMC phase. As far as we know, this result has not yet been measured in TI experiments. These observables are also interesting to see the magnetic properties of materials. We hope to see the measurement near future and compare it to our result.

One of important issues of this paper is to identify the impurity and to clarify whether this impurity actually plays a role of magnetic impurity. The experimental data show the phase transition with increasing obvious magnetic impurity. To our knowledge, there is no holographic construction for a magnetic impurity relevant to our case. However, the axion parameter really changes the susceptibility and the magnetization with fixed the other external sources. This guarantees that the linear-axion strength $\beta$ affects the magnetic moment of the dual system, as well as it describes a usual impurity.

Also, due to the external magnetic field, TRS is broken. So we need to know the time reversal transformation of each field including the linear-axion. In order to figure out this issue in a symmetry perspective, the time dependent fluctuation of black brane backgrounds should be studied. However, in this note we only consider DC transport coefficients obtained by static fluctuations (\ref{feq1}-\ref{feq3}). One of our future project is to obtain AC transport coefficients. To find the quantities, the time-dependent fluctuation should be investigated. So we leave analysis of TRS properties of the linear-axion as a future study.

In our consideration we didn't turn on the source $\mathcal{J}$ of the real scalar operator $\mathcal{O}$ to simplify the problem. However an extension using this source is another interesting subject. It can be regarded as another kind of impurity density. This is very similar to hysteric conductivities measured in a different kind of topological insulators that show hysteric conductivities as soon as the source is turned on \cite{Cr-dopedTI-1,Cr-dopedTI-2,Cr-dopedTI-3}.

In addition we consider only the phase transition between the WL phase and the HMC phase. In a preliminary study which is not revealed in this paper, we found that the full solution space has more various types of numerical solutions. So we leave further analysis on the full solution space as one possible future study. Other extensions with additional matter fields are also possible and interesting to understand underlying physics of this hysteric phase transition. We hope that these generalizations of the present model could describe the corssover and the phase transition among all the phases (WAL, WL and HMC phases) in the near future.

\section*{Appendix}

\section*{A. Magnetization from Scaling Symmetry Technique}

In this section we introduce another way to obtain the magnetization. This will confirm the magnetization obtained by consideration of the heat current in section \ref{DCconductivity}. In order to get the magnetization, we use a scaling symmetry technique developed in \cite{Banados:2005hm,Ahn:2015uza,Hyun:2015tia,Ahn:2015shg,Hyun:2017nkb,Kim:2019lxb}. This technique is based on a scaling symmetry in a reduced action. So we start with the reduced action using (\ref{Ansatz00}). Then the bulk action (\ref{S_B}) can be written as 
\begin{align}
S_B = \int dr d^3 x L_{red} = \int dr d^3 s \left(L_0 + L_{bs}\right)~,
\end{align}
where $L_0$ and $L_{bs}$ are given as follows:
\begin{align}
L_0 =& \frac{r^2 e^{W(\infty )-W} A_t'^2}{32 \pi  G L^2}+\frac{e^{W-W(\infty )}}{8 \pi  G L^4}\left(3 r^2-\frac{1}{4} L^2 m^2 r^2 \phi^2-\frac{1}{4} r^2 U \phi'^2-r\, U'-U\right)\\L_{bs}=&-\frac{e^{W-W(\infty )}}{32 \pi  G}\left( 2 \beta ^2 +\frac{\mathcal{H}^2 L^4}{r^2}\right)+\frac{  \mathcal{H} L  A_t'  \eta\,  \phi}{8 \pi  G }~.
\end{align}
Here we discarded some total derivative terms which doesn't spoil the following consideration.

Now let us assign the following transformation to each field:
\begin{align}
&\delta_0 U = \sigma  \left(2 U-r U'\right),~\delta_0 e^W =\sigma\left(-2 e^W -r (e^W)' \right),\nonumber\\&\delta_0\phi = \sigma \left(-r\phi' \right),~\delta_0 A_t = \sigma\left( -2 A_t - rA_t' \right),  
\end{align}  
where $\sigma$ is a small parameter. $L_0$ is invariant under the scaling transformation $\delta_0$ up to total derivative as follows:
\begin{align}
\delta_0 L_0 = \left(- r L_0 \right)'~,
\end{align}
where we used the equations of motion for $W(r)$ from the action $L_0$. Together with $\delta_0$, one can add the following parameter transformation:
\begin{align}
\delta_c H = 2 \sigma H~,~\delta_c \beta = \sigma \beta~.
\end{align}
The above transformation compensates the scaling invariance of the full reduced action, i.e.
\begin{align}
\left(\delta_0 + \delta_c\right)L_{red} = \left( - r L_{red} \right)'~.
\end{align}
Then this transformation is an approximated symmetry and the violation can be compensated by the parameter variation $\delta_c$.

Using the equations of motion for the fields, we arrive at a partially conserved charge (PCC) equation as follows: 
\begin{align}\label{PConservation}
\mathcal{C}' =- \frac{1}{2}\delta_c L_{red}~,
\end{align}
where the PCC $\mathcal{C}$ is defined by
\begin{align}
\mathcal{C}\equiv \frac{1}{2}\left( \frac{\delta_0 L_{red}}{\delta_0 \Psi'} \delta_0\Psi + r L_{red} \right). 
\end{align}
$\Psi$ denotes the collective field of $\left\{ U, e^W, \phi, A_t \right\}$. The explicit form of PCC in terms of the ansatz (\ref{Ansatz00}) is 
\begin{align}\label{PCC}
\mathcal{C} = \frac{e^{W-W(\infty )} \left(r^3 U \phi'^2+2 r^2 U'-4 r U\right)}{32 \pi  G L^4}-\frac{L q A_t}{16 \pi  G}.
\end{align}
The PCC equation (\ref{PConservation}) is a consequence of the broken scaling symmetry by $L_{bs}$.

Now, (\ref{PConservation}) can be integrated over from $r_h$ to UV cut-off $\Lambda$.
\begin{align}\label{Smarr00}
\mathcal{C}(\Lambda) +\frac{1}{2} \int_{r_h}^{\Lambda} dr\delta_c L_{red}  =\mathcal{C}(r_h) ~,
\end{align}
where $\mathcal{C}(r_h)$ turns out to be $s\, T$. The asymptotic value of PCC (\ref{PCC}) is 
\begin{align}
\mathcal{C}(\Lambda) = \frac{\tilde{\Lambda } r_h^3 \tilde{\beta }^2 }{16 \pi  G L^4} +\frac{r_h^3 \left(3 \tilde{M}-2 \tilde{J} \tilde{\mathcal{O}}\right)}{16 \pi  G L^4} -\frac{\mu  \tilde{q} r_h^2}{16 \pi  G L^2} + \mathcal{O}\left( \frac{1}{\tilde \Lambda} \right)~,
\end{align}
where $\tilde\Lambda = \Lambda/r_h$. Using all the above expressions and the asymptotic behavior (\ref{asymptotics}) of fields with $\tilde{J}=0$, it turns out that (\ref{Smarr00}) gives us the thermodynamic relation
\begin{align}
\mathcal{E} + \mathcal P = \mu \mathcal{Q} + s T~~,
\end{align}
where $\mathcal{E}$, $\mathcal{P}$ and $\mathcal{Q}$ are the energy density, the pressure and the charge density of the dual field theory, respectively. They can be evaluated by holographic renormalization with the counter terms:
\begin{align}
S_{ct} =- \frac{1}{16\pi G}\int_{r=\Lambda}d^3 x \sqrt{-\gamma} \left( \frac{4}{L} + \frac{\phi^2}{2L} - \frac{L}{2}\left(\nabla\psi^{\mathcal I}\cdot\nabla\psi^{\mathcal I}  \right) \right)~,
\end{align} 
where $\gamma$ is the determinant of the induced metric $\gamma_{\mu\nu}$ from the ADM decomposition ($ds^2=\gamma_{\mu\nu}+ N^2 dr^2$).

More explicitly, the holographic energy density $\mathcal{E}=T^{tt}$ given by
\begin{align}
T^{tt} =  \frac{r_h^3 \tilde{M}}{8\pi G L^4}.
\end{align} 
from the boundary energy-momentum tensor $T^{\mu\nu}$. In addition the holographic charge density $\mathcal Q$ is defined by $\mathcal{Q} \equiv \frac{r_h^2\tilde{q}}{16\pi G L^2}$. See (\ref{Current}) for the expression of the holographic current whose time-component is equivalent to $\mathcal{Q}$. Since the system is a homogeneous system, the pressure is nothing but the minus Euclidean on-shell action density $\mathcal{W}$, i.e. $\mathcal{P} =-\mathcal{W}$. The pressure consists of three parts :
\begin{align}
\mathcal{P} = \mathcal{P}_{int} + \beta \mathcal{M}_\beta + \mathcal{H} \mathcal{M}_\mathcal{H} ~.
\end{align}  
Each term is given as follows :
\begin{align}
\mathcal{P}_{int} =& T^{xx} = \frac{r_h^3 \tilde{M}}{16\pi G L^4}\\
\mathcal{M}_\beta =&  \frac{\tilde{\beta}\,r_h^2}{16\pi G L^2}\left[1-\int_1^{\tilde \Lambda} d\tilde{r} \left( e^{w(\tilde r)-w(\tilde\Lambda)} -1 \right)\right]\\
\mathcal{M}_\mathcal{H} =& \frac{  r_h }{8 \pi  G } \int_1^{\tilde\Lambda} d\tilde{r}\, e^{w\left(\tilde{r}\right)-w\left(\tilde{\Lambda }\right)}\left\{ \tilde{q} \left(\frac{\eta  \varphi \left(\tilde{r}\right)}{\tilde{r}^2}\right)  -\tilde{H} \left(\frac{4 \eta ^2 \varphi \left(\tilde{r}\right)^2+1}{2 \tilde{r}^2}\right)\right\},\label{eq:magnetization}
\end{align}
where $\mathcal{P}_{int}$ can be regarded as the internal pressure. In addition $\mathcal{M}_\beta$ is the dual thermodynamic variable to $\beta$. Also, $\mathcal{M}_{\mathcal H}$ denotes the magnetization which is thermodynamic dual to the external magnetic field $\mathcal H$. This expression is the same form of magnetization from the heat current (\ref{Heat Current}). Therefore this magnetization from the magnetic pressure provides an independent consistency check for the magnetization included in the heat current.

\section*{Acknowledgments}
K. K. Kim was supported by Basic Science Research Program through the National Research Foundation of Korea(NRF)  grant No.  NRF-2019R1A2C1007396. K.-Y. Kim was supported by NRF funded by the Ministry of Science, ICT $\&$ Future Planning (NRF- 2021R1A2C1006791) and the GIST Research Institute(GRI) grant funded by the GIST in 2021. S. J. Sin was supported by Mid-career Researcher Program through the NRF grant No. NRF-2016R1A2B3007687. The work of YS was supported by Basic Science Research Program through NRF grant No. NRF-2019R1I1A1A01057998.
  K.K.Kim acknowledges the hospitality at APCTP where a part of this work was done.



\begin{thebibliography}{99}


\bibitem{HasanKane}
M.~Z~Hasan and C.~L.~Kane,
``Colloquium: Topological insulators,''
Rev.\ Mod.\ Phys.\ {\bf 82}, 3045(2010)


\bibitem{Ando:2013bqa} 
  Y.~Ando,
  ``Topological Insulator Materials,''
  J.\ Phys.\ Soc.\ Jap.\  {\bf 82}, no. 10, 102001 (2013)
  [arXiv:1304.5693 [cond-mat.mtrl-sci]].

 
\bibitem{TIexps-1}

L.~Bao,  W.~Wang,  N.~Meyer , Y.~Liu, C.~Zhang, K.~Wang, P.~Ai and F.~Xiu,
``Quantum Corrections Crossover and Ferromagnetism in Magnetic Topological Insulators,'' 
Scientific~Reports\ {\bf 3}, 2391 (2013) 

\bibitem{TIexps-2}

Y.~Ni, Z.~Zhang, I.~C.~Nlebedim, R.~L.~Hadimani, G.~Tuttle, and D.~ C.~Jiles,
``Ferromagnetism of magnetically doped topological insulators in Cr${}_\text{x}$Bi${}_\text{2-x}$ Te${}_\text{3}$,''
Journal of Applied Physics {\bf 117}, 17C748 (2015)





\bibitem{Cr-dopedTI-1} 
X.~Kou, M.~Lang, Y.~Fan, Y.~Jiang, T.~Nie, J.~Zhang, W.~Jiang, Y.~Wang, Y.~Yao, L.~ He, and K.~ L.~ Wang,
``Interplay between Different Magnetisms in Cr-Doped Topological Insulators,'' ACS Nano (2013), {\bf 7}, 10, 9205–9212.


\bibitem{Cr-dopedTI-2}
J.~Zhang, C.~Chang, P.~Tang, Z.~Zhang, X.~Feng, K.~Li, L.~Wang, X.~ Chen, C.~Liu, W.~Duan, K.~He, Q.~Xue, X.~Ma and Y.~Wang,
``Topology-Driven Magnetic Quantum Phase Transition in Topological Insulators,''
Science\ {\bf 339}, 1582 (2013) 


\bibitem{Cr-dopedTI-3}
X.~Kou, Y.~Fan,M.~Lang, P.~Upadhyaya, K.~L.~Wang,
``Magnetic  topological insulators and quantum anomalous hall effect,''
Solid\ State\ Communications\ , {\bf 215-216}, 34(2015)


\bibitem{WAL-WL}


F.~V.~Tikhonenko, A.~ A.~ Kozikov, A.~ K.~ Savchenko and R.~V.~ Gorbachev, ``Transition between Electron Localization and Antilocalization in Graphene'' PRL {\bf 103}, 226801 (2009)



H.~ Lu, J.~ Shi, and S.~ Shen, ``Competition between Weak Localization and Antilocalization in Topological Surface States'' PRL {\bf 107}, 076801 (2011)


M.~ Liu, J.~ Zhang, C.~ Chang, Z.~ Zhang, X.~ Feng, K.~ Li, K.~ He, L.~ Wang,
X.~ Chen, X.~ Dai, Z.~ Fang, Q.~ Xue, X.~ Ma and Y.~ Wang, ``Crossover between Weak Antilocalization andWeak Localization in a Magnetically Doped Topological Insulator'' PRL {\bf 108}, 036805 (2012)



\bibitem{Hikami_model} 
S.~Hikami, A.~I.~Larkin and Y.~Nagaoka,
``Spin-Orbit Interaction and Magnetoresistance in the Two Dimensional Random System,'' Prog.\ Theor.\ Phys.\ {\bf 63}, 707 (1980).






 


\bibitem{YYYang} 
Yan-Yan~Yang,~Ming-Xun~Deng,~Wei~Luo,~R.~Ma,~Shi-Han~Zheng and Rui-Qiang~Wang,
  ``Anomalous Nernst effect on a magnetically-doped topological insulator surface: A Green's function approach,''
  Phys.\ Rev.\ B {\bf 98}, 235152 (2018)

\bibitem{Zarezad} 
A.~N.~Zarezad  and J.~Abouie
  ``Transport in magnetically doped topological insulators: Effects of magnetic clusters,''
  Phys.\ Rev.\ B {\bf 98}, 155413 (2018)

 



\bibitem{Seo:2017oyh} 
  Y.~Seo, G.~Song and S.~J.~Sin,
  ``Strong Correlation Effects on Surfaces of Topological Insulators via Holography,''
  Phys.\ Rev.\ B {\bf 96}, no. 4, 041104 (2017)
  [arXiv:1703.07361 [hep-th]].
  
\bibitem{Seo:2017yux}
Y.~Seo, G.~Song, C.~Park and S.~J.~Sin,
``Small Fermi Surfaces and Strong Correlation Effects in Dirac Materials with Holography,''
JHEP \textbf{10}, 204 (2017)
[arXiv:1708.02257 [hep-th]].


\bibitem{Maldacena:1997re} 
  J.~M.~Maldacena,
  ``The Large N limit of superconformal field theories and supergravity,''
  Int.\ J.\ Theor.\ Phys.\  {\bf 38}, 1113 (1999)
  [Adv.\ Theor.\ Math.\ Phys.\  {\bf 2}, 231 (1998)]
  [hep-th/9711200].


\bibitem{Aharony:1999ti} 
  O.~Aharony, S.~S.~Gubser, J.~M.~Maldacena, H.~Ooguri and Y.~Oz,
  ``Large N field theories, string theory and gravity,''
  Phys.\ Rept.\  {\bf 323}, 183 (2000)
  [hep-th/9905111].


\bibitem{Hartnoll:2008vx} 
  S.~A.~Hartnoll, C.~P.~Herzog and G.~T.~Horowitz,
  ``Building a Holographic Superconductor,''
  Phys.\ Rev.\ Lett.\  {\bf 101}, 031601 (2008)
  [arXiv:0803.3295 [hep-th]].


\bibitem{Hartnoll:2007ih} 
  S.~A.~Hartnoll, P.~K.~Kovtun, M.~Muller and S.~Sachdev,
  ``Theory of the Nernst effect near quantum phase transitions in condensed matter, and in dyonic black holes,''
  Phys.\ Rev.\ B {\bf 76}, 144502 (2007)
  [arXiv:0706.3215 [cond-mat.str-el]].


\bibitem{Hartnoll:2009sz} 
  S.~A.~Hartnoll,
  ``Lectures on holographic methods for condensed matter physics,''
  Class.\ Quant.\ Grav.\  {\bf 26}, 224002 (2009)
  [arXiv:0903.3246 [hep-th]].


\bibitem{Hartnoll:2009ns} 
  S.~A.~Hartnoll, J.~Polchinski, E.~Silverstein and D.~Tong,
  ``Towards strange metallic holography,''
  JHEP {\bf 1004}, 120 (2010)
  [arXiv:0912.1061 [hep-th]].


\bibitem{Hartnoll:2007ip} 
  S.~A.~Hartnoll and C.~P.~Herzog,
  ``Ohm's Law at strong coupling: S duality and the cyclotron resonance,''
  Phys.\ Rev.\ D {\bf 76}, 106012 (2007)
  [arXiv:0706.3228 [hep-th]].

\bibitem{Lindgren:2015lia}
J.~Lindgren, I.~Papadimitriou, A.~Taliotis and J.~Vanhoof,
``Holographic Hall conductivities from dyonic backgrounds,''
JHEP \textbf{07}, 094 (2015)
[arXiv:1505.04131 [hep-th]].

\bibitem{Lindgren:2015lia2}
M.~M.~Caldarelli, A.~Christodoulou, I.~Papadimitriou and K.~Skenderis,
``Phases of planar AdS black holes with axionic charge,''
JHEP \textbf{04}, 001 (2017)
[arXiv:1612.07214 [hep-th]].

\bibitem{Kim:2019lxb} 
  K.~K.~Kim, K.~Y.~Kim, Y.~Seo and S.~J.~Sin,
  ``Building Magnetic Hysteresis in Holography,''
  JHEP {\bf 1907}, 158 (2019)
  [arXiv:1902.10929 [hep-th]].


\bibitem{Andrade:2013gsa} 
  T.~Andrade and B.~Withers,
  ``A simple holographic model of momentum relaxation,''
  JHEP {\bf 1405}, 101 (2014)
  [arXiv:1311.5157 [hep-th]].


\bibitem{Gouteraux:2014hca} 
  B.~Goutéraux,
  ``Charge transport in holography with momentum dissipation,''
  JHEP {\bf 1404}, 181 (2014)
  [arXiv:1401.5436 [hep-th]].


\bibitem{Donos:2014cya} 
  A.~Donos and J.~P.~Gauntlett,
  ``Thermoelectric DC conductivities from black hole horizons,''
  JHEP {\bf 1411}, 081 (2014)
  [arXiv:1406.4742 [hep-th]].


\bibitem{Donos:2014uba} 
  A.~Donos and J.~P.~Gauntlett,
  ``Novel metals and insulators from holography,''
  JHEP {\bf 1406}, 007 (2014)
  [arXiv:1401.5077 [hep-th]].


\bibitem{Kim:2014bza} 
  K.~Y.~Kim, K.~K.~Kim, Y.~Seo and S.~J.~Sin,
  ``Coherent/incoherent metal transition in a holographic model,''
  JHEP {\bf 1412}, 170 (2014)
  [arXiv:1409.8346 [hep-th]].


\bibitem{Blake:2015ina} 
  M.~Blake, A.~Donos and N.~Lohitsiri,
  ``Magnetothermoelectric Response from Holography,''
  JHEP {\bf 1508}, 124 (2015)
  [arXiv:1502.03789 [hep-th]].


\bibitem{Kim:2015wba} 
  K.~Y.~Kim, K.~K.~Kim, Y.~Seo and S.~J.~Sin,
  ``Thermoelectric Conductivities at Finite Magnetic Field and the Nernst Effect,''
  JHEP {\bf 1507}, 027 (2015)
  [arXiv:1502.05386 [hep-th]].


\bibitem{Kim:2015dna} 
  K.~Y.~Kim, K.~K.~Kim and M.~Park,
  ``A Simple Holographic Superconductor with Momentum Relaxation,''
  JHEP {\bf 1504}, 152 (2015)
  [arXiv:1501.00446 [hep-th]].


\bibitem{Kim:2016hzi} 
  K.~K.~Kim, M.~Park and K.~Y.~Kim,
  ``Ward identity and Homes’ law in a holographic superconductor with momentum relaxation,''
  JHEP {\bf 1610}, 041 (2016)
  [arXiv:1604.06205 [hep-th]].


\bibitem{Donos:2015bxe} 
  A.~Donos, J.~P.~Gauntlett, T.~Griffin and L.~Melgar,
  ``DC Conductivity of Magnetised Holographic Matter,''
  JHEP {\bf 1601}, 113 (2016)
  [arXiv:1511.00713 [hep-th]].


\bibitem{Henningson:1998gx} 
  M.~Henningson and K.~Skenderis,
  ``The Holographic Weyl anomaly,''
  JHEP {\bf 9807}, 023 (1998)
  [hep-th/9806087].


\bibitem{deHaro:2000vlm} 
  S.~de Haro, S.~N.~Solodukhin and K.~Skenderis,
  ``Holographic reconstruction of space-time and renormalization in the AdS / CFT correspondence,''
  Commun.\ Math.\ Phys.\  {\bf 217}, 595 (2001)
  [hep-th/0002230].


\bibitem{Skenderis:2002wp} 
  K.~Skenderis,
  ``Lecture notes on holographic renormalization,''
  Class.\ Quant.\ Grav.\  {\bf 19}, 5849 (2002)
  [hep-th/0209067].


\bibitem{Seo:2015pug} 
  Y.~Seo, K.~Y.~Kim, K.~K.~Kim and S.~J.~Sin,
  ``Character of matter in holography: Spin–orbit interaction,''
  Phys.\ Lett.\ B {\bf 759}, 104 (2016)
  [arXiv:1512.08916 [hep-th]].


\bibitem{Seo:2016vks} 
  Y.~Seo, G.~Song, P.~Kim, S.~Sachdev and S.~J.~Sin,
  ``Holography of the Dirac Fluid in Graphene with two currents,''
  Phys.\ Rev.\ Lett.\  {\bf 118}, no. 3, 036601 (2017)
  [arXiv:1609.03582 [hep-th]].


\bibitem{Banados:2005hm} 
  M.~Banados and S.~Theisen,
  ``Scale invariant hairy black holes,''
  Phys.\ Rev.\ D {\bf 72}, 064019 (2005)
  [hep-th/0506025].


\bibitem{Ahn:2015uza} 
  B.~Ahn, S.~Hyun, S.~A.~Park and S.~H.~Yi,
  ``Scaling symmetry and scalar hairy rotating AdS$_3$ black holes,''
  Phys.\ Rev.\ D {\bf 93}, no. 2, 024041 (2016)
  [arXiv:1508.06484 [hep-th]].


\bibitem{Hyun:2015tia} 
  S.~Hyun, J.~Jeong, S.~A.~Park and S.~H.~Yi,
  ``Scaling symmetry and scalar hairy Lifshitz black holes,''
  JHEP {\bf 1510}, 105 (2015)
  [arXiv:1507.03574 [hep-th]].


\bibitem{Ahn:2015shg} 
  B.~Ahn, S.~Hyun, K.~K.~Kim, S.~A.~Park and S.~H.~Yi,
  ``Holography without counter terms,''
  Phys.\ Rev.\ D {\bf 94}, no. 2, 024043 (2016)
  [arXiv:1512.09319 [hep-th]].


\bibitem{Hyun:2017nkb} 
  S.~Hyun, J.~Jeong, S.~A.~Park and S.~H.~Yi,
  ``Thermodynamic Volume and the Extended Smarr Relation,''
  JHEP {\bf 1704}, 048 (2017)
  [arXiv:1702.06629 [hep-th]].





\end{thebibliography}
\end{document}